\documentclass[preprint,superscriptaddress,nofootinbib]{revtex4-1}
\usepackage{graphicx}
\usepackage{amsmath,amssymb,color,colortbl}

\usepackage{dcolumn}
\usepackage{bm}

\usepackage{epsfig}    
\usepackage{ulem}

\newcommand{\nn}{\nonumber}
\newcommand{\be}{\begin{eqnarray}}
\newcommand{\ee}{\end{eqnarray}}
\catcode`\@=11

\def\lsim{\mathrel{\mathpalette\@versim<}}
\def\gsim{\mathrel{\mathpalette\@versim>}}
\def\@versim#1#2{\vcenter{\offinterlineskip
\ialign{$\m@th#1\hfil##\hfil$\crcr#2\crcr\sim\crcr } }}
\catcode`\@=12

\allowdisplaybreaks
\def\thefootnote{\fnsymbol{footnote}}

\begin{document}
\title{
Multicomponent Dark Matter in Radiative Seesaw Models 
}

\author{Mayumi \surname{Aoki}}
\email{mayumi@hep.s.kanazawa-u.ac.jp}
\affiliation{Institute for Theoretical Physics, Kanazawa University, Kanazawa 920-1192, Japan}

\author{Daiki \surname{Kaneko}}
\email{d$_$kaneko@hep.s.kanazawa-u.ac.jp}
\affiliation{Institute for Theoretical Physics, Kanazawa University, Kanazawa 920-1192, Japan}

\author{Jisuke \surname{Kubo}}
\email{jik@hep.s.kanazawa-u.ac.jp}
\affiliation{Institute for Theoretical Physics, Kanazawa University, Kanazawa 920-1192, Japan}

\preprint{KANAZAWA-17-10}

\begin{abstract}
We discuss radiative seesaw models, in which
an exact $Z_2\times Z_2'$ symmetry is imposed.
Due to the exact $Z_2\times Z_2'$ symmetry,
neutrino masses are generated at a two-loop level and at least two extra stable electrically 
neutral particles are
predicted.
We consider two models: one has a multi-component dark matter system and the other one has
a dark radiation in addition to a dark matter. 
In the multi-component dark matter system,
non-standard dark matter annihilation processes exist. 
We find 
that they play important roles in determining the relic abundance
and also responsible for the monochromatic neutrino lines resulting from 
the dark matter annihilation process.
In the model with the dark radiation, 
the structure of the Yukawa coupling is considerably constrained and 
gives an interesting
relationship among
cosmology, lepton flavor violating decay of the charged leptons
and the decay of the inert Higgs bosons.
 \end{abstract}
 
\setcounter{footnote}{0}
\def\thefootnote{\arabic{footnote}}
\maketitle
%
 
\clearpage
\section{Introduction}
Neutrino oscillation experiments show that the neutrinos have tiny masses and mix with each other.
It is a clear evidence for physics beyond the standard model (SM), since the SM has no mechanism for giving masses to the neutrinos. 
The global fit \cite{Esteban:2016qun} shows that the mass-squared differences of the 
neutrinos are
 $ 
  \Delta m_{21}^2 = 7.50^{+0.19}_{-0.17}\times 10^{-5}$ eV$^2$
and
$\Delta m_{31}^2 =2.524^{+0.039}_{-0.040}
~(-2.514^{+0.038}_{-0.041})\times 10^{-3}\mbox{ eV}^2~$
    for normal (inverted) mass hierarchy.  
    The cosmological data, on the other hand, gives the upper bound of the sum of the neutrino masses as $\Sigma_j m_{\nu_j} < 0.23$ eV~\cite{Ade:2015xua}, a scale twelve  orders of magnitudes smaller than
the electroweak scale.
It is one of the most important problems of particle physics to reveal the origin of the tiny masses for the neutrinos.

Type-I seesaw mechanism \cite{Minkowski:1977sc}
is one of the attractive way to realize the tiny masses of the neutrinos, where
the right-handed neutrinos are introduced to the SM.
If the neutrino Yukawa coupling for the Dirac neutrino mass is ${\cal O}(1)$,
the mass of the right-handed neutrino has to be around ${\cal O}$($10^{15}$) GeV
to obtain eV-scale neutrinos.
The mass scale of ${\cal O}$($10^{15}$) GeV is obviously beyond the reach of collider experiments. Even  
for the mass of the right-handed neutrinos around ${\cal O}$(1) TeV,
the direct search of the right-handed neutrinos would be difficult because of the tiny neutrino Yukawa couplings of ${\cal O}(10^{-6})$.
 
Another attractive way to give the neutrino masses is a radiative generation 
(the so-called radiative seesaw model). 
The original idea of radiatively generating neutrino masses due to TeV-scale physics has been proposed by 
Zee~\cite{Zee:1980ai}, in which the neutrino masses are induced at the one-loop level 
because of the addition of
an isospin doublet scalar field 
and a charged singlet field to the SM.
Another possibility for generating neutrino masses via the new scalar particles is {\it e.g.} the Zee-Babu model~\cite{Zee:1985id}, in which the neutrino masses arise at the two-loop level. 

A further extension with a TeV-scale right-handed neutrino has been proposed in Ref.~\cite{Krauss:2002px}. 
In this model
the neutrino masses are induced at the three-loop level, 
where the Dirac neutrino mass at the tree level
is forbidden due to an exact $Z_2$ symmetry.
The right-handed neutrino is odd under the $Z_2$ and becomes a candidate of dark matter (DM). 
The idea of simultaneous explanation for the neutrino masses via the radiative seesaw mechanism and 
the stability of DM by introducing 
an exact discrete symmetry 
has been discussed in many models (see, {\it e.g.}, Refs.~\cite{
 Ma:2006km,
 Nasri:2001ax,
Ma:2007yx,
Kajiyama:2013rla,
Baek:2013fsa,
 Aoki:2013gzs,
Aoki:2014lha,
Ma:2006uv}
and the recent review~\cite{Cai:2017jrq} and references therein).

The model proposed by Ma in Ref.~\cite{Ma:2006km} is one of the simplest radiative seesaw model with DM candidates.
The model has the $Z_2$-odd right-handed neutrinos $N_k$ and the inert doublet scalar field $\eta=(\eta^+, \eta^0_R+i \eta^0_I)^T$.
The neutrino masses are generated at the one-loop level, in which the Yukawa interactions 
$Y^\nu_{ik}L_i\eta N_k$ and the scalar interaction 
$(\lambda_5/2) (H^\dag \eta )^2$ contribute to the neutrino mass generation.
The mass matrix is expressed as  
\begin{align}
({\cal M}_\nu)_{ij} &= \sum_k 
\frac{Y^\nu_{ik} Y^\nu_{jk} M_k}{32 \pi^2 }
\left[\frac{m_{\eta_R^0}^2}{m_{\eta_R^0}^2-M_k^2}\ln\left(\frac{m_{\eta_R^0}}{M_k}\right)^2
-\frac{m_{\eta_I^0}^2}{m_{\eta_I^0}^2-M_k^2}\ln\left(\frac{m_{\eta_I^0}}{M_k}\right)^2
\right] , \nonumber
\end{align}
where $M_k$ is the Majorana mass of the $k$-th generation of right-handed neutrino, $m_{\eta_R^0}$ and $m_{\eta_I^0}$ are
the mass of the $\eta^0_R$ and $\eta^0_I$, respectively.
In this model, we have two scenarios with respect 
to the DM candidate; the lightest right-handed neutrino $N_1$ or the lighter $Z_2$-odd neutral scalar field ($\eta^0_R$ or $\eta^0_I$).
The phenomenology of the model is studied in Refs.~\cite{Kubo:2006yx,Bouchand:2012dx, Ma:2001mr, Suematsu:2009ww,Sierra:2008wj,
Gelmini:2009xd}.

A DM candidate can be made stable by an  unbroken symmetry. The simplest possibility of such a symmetry is 
$Z_2$ symmetry as in the above models.
However, if the DM stabilizing symmetry is larger than $Z_2$: $Z_N$ $(N \ge 4)$ or a product of two or more $Z_2$’s,
the DM is consisting of stable multi-DM particles 
(multicomponent DM system). 
A supersymmetric extension of the radiative seesaw model of Ref.~\cite{Ma:2006km} is an example~\cite{Ma:2006uv}.
Other possibilities with multicomponent DM are widely discussed in~\cite{Ma:2007yx,Kajiyama:2013rla,Baek:2013fsa,Aoki:2013gzs,Aoki:2014lha,Bhattacharya:2016ysw,Berezhiani:1990sy,Aoki:2012ub}.

In this paper we study two models of the two-loop extension of the model by Ma~\cite{Ma:2006km},
we call them as ``model A" and ``model B", 
in which due to the $Z_2 \times Z_2'$ symmetry a set of stable particles can exist.
Introducing two new scalar fields, the $\lambda_5$ term is generated radiatively in the model A~\cite{Aoki:2013gzs,Aoki:2014lha}.
In this model we discuss the three component DM system in which the two new scalar fields and a right-handed neutrino are 
the DM candidate. Such case has been discussed in~\cite{Aoki:2014lha}, however, we reanalyze the model
since the benchmark points studied in~\cite{Aoki:2014lha}, where the masses of both new scalars are several hundred GeV,
 has been excluded by the recent results of the direct detection DM experiments.
In this paper we focus on the case where the mass of one of the scalar DMs is close to the half of the Higgs boson mass to satisfy the constraints from the direct detection.
In the model B the right-handed neutrinos have the mass radiatively generated through the one loop of internal
new fermion and scalar fields. 
We identify the lightest right-handed neutrino as dark radiation.

We start in section II by writing down a set of the Boltzmann equations of the multicomponent DM system.
The model A is discussed in section III by following Ref.~\cite{Aoki:2014lha}. 
In section IV we discuss the model B 
and relate dark radiation with the lepton flavor violating decay of the charged leptons and the decay of the inert Higgs bosons.
Summery and discussion are given in section V.

\section{Multicomponent dark matter systems}
In the case of one-component DM the relic density of DM $\chi$ is determined by  the  Boltzmann equation
\begin{align}
\dot{n}_\chi + 3Hn_\chi
=&
-\langle \sigma_{\chi\chi \rightarrow XX'} v \rangle (n_\chi^2- \bar{n}_\chi^2 ) \,,
\label{eq:boltz1}
\end{align}
where $n_\chi$ is the DM number density, $\bar n_\chi$ is the equilibrium number density
and $\langle \sigma_{\chi\chi \rightarrow XX'} v \rangle$ is the thermally averaged cross section
for $\chi\chi \rightarrow XX'$. Here $X$ and $X'$ stand for the SM particles.  
The Hubble parameter $H$ is given by $H=1.66\times g_*^{1/2} T^2/M_{\rm PL}$,
where 
$g_*$ is the total number of effective
degrees of freedom,  $T$ and $M_{\rm PL}$ are the temperature
and the Planck mass, respectively.
It is convenient to rewrite the equation in terms of dimensionless quantities; 
the number per comoving volume $Y_\chi=n_\chi/s$ and the inverse temperature $x=m/T$. 
Here $s$ is the entropy density $s=(2 \pi^2/45 ) g_* T^3$ and $m$ is the mass of the DM particle.
Using the replacement of $dx/dt=H|_{T=m}/x$, we obtain
\begin{align}
\frac{d Y_\chi}{d x}&=-0.264~ g_*^{1/2} \frac{m M_{\rm PL}}{x^2} 
~\langle \sigma_{\chi\chi \rightarrow XX'} v \rangle \left(  Y_\chi Y_\chi -\bar{Y}_{\chi}\bar{Y}_{\chi}\right) \,.
\end{align}
The thermally averaged cross section $\langle \sigma_{\chi\chi \rightarrow XX'} v \rangle$ of ${\cal O}(10^{-9})$ GeV with a DM mass of 100 GeV gives $Y_\chi \simeq 10^{-12}$, which is consistent with the observed value of the relic abundance $\Omega h^2\simeq 0.12$~\cite{Ade:2013zuv}.

\begin{figure}[t]
\begin{center}
  \includegraphics[width=13cm]{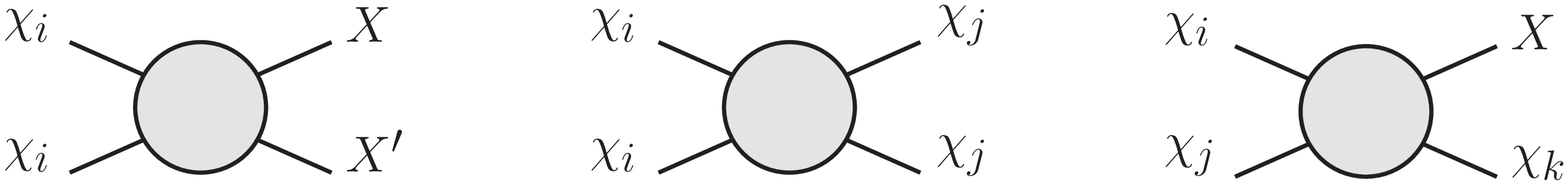}
\end{center}
\caption{\label{NS2} 
Standard annihilation (left),
DM conversion (middle) and semi-annihilation (right).
}
\end{figure}

In the multicomponent DM system three types of processes enter in the Boltzmann equations
\footnote{Semi-annihilation processes also exist in one-component DM systems
when DM is a $Z_3$ charged particle \cite{D'Eramo:2010ep} or a vector boson \cite{Hambye:2008bq}.}:
\begin{align}
&\chi_i\chi_i \leftrightarrow XX'~~~~ \mbox{(standard annihilation)}, \label{p0}\\
 &  \chi_i\chi_i   \leftrightarrow  \chi_j\chi_j ~~~~ \mbox{(DM conversion)},         \label{p1}\\
& \chi_i\chi_j \leftrightarrow \chi_kX_{} ~~~~ \mbox{(semi-annihilation)}\,.
\label{p2}
\end{align}
See Fig.~\ref{NS2} for a depiction of three types of processes.
Here we assume that
none of the DM particles have the same quantum number with respect to the DM stabilizing symmetry. 
The Boltzmann equations for the DM particle $\chi_i$ with mass $m_i$
are
\begin{align}
\frac{d Y_i}{d x}
&=-0.264~ g_*^{1/2} \frac{\mu M_{\rm PL}}{x^2} 
\bigg\{
\langle \sigma_{\chi_i\chi_i \to XX'} v \rangle
\left(  Y_i Y _i-\bar{Y}_{i}\bar{Y}_{i}\right)
\frac{}{}
\nn\\
 &\left. +
\sum_{i >j }
\langle \sigma_{\chi_i\chi_i   \to  \chi_j\chi_j}v\rangle
\!\!\left(  Y_i Y _i-\frac{Y_j Y_j}
{\bar{Y}_{j}\bar{Y}_{j}} \bar{Y}_{i}\bar{Y}_{i}
\right)
\!-\!\sum_{j >i}
\langle \sigma_{\chi_j\chi_j  \to  \chi_i\chi_i} v \rangle
\left(  Y_j Y _j-\frac{Y_i Y_i}
{\bar{Y}_{i}\bar{Y}_{i}} \bar{Y}_{j}\bar{Y}_{j}
\right)~\right.~\nn\\
 &\left.
 +\!\sum_{j,k}
 \langle \sigma_{\chi_i\chi_i   \to  \chi_kX_{ijk}} v \rangle 
\left(  Y_i Y _j-\frac{Y_k}
{\bar{Y}_{k}} \bar{Y}_{i}\bar{Y}_{j}
\right)-\!\sum_{j,k}
\langle \sigma_{\chi_j\chi_k   \to \chi_kX_{ijk}} v \rangle
\left(  Y_j Y _k-\frac{Y_i}
{\bar{Y}_{i}} \bar{Y}_{j}\bar{Y}_{k}
\right)
\right\}
~.
\label{boltz}
\end{align}
Here 
$x=\mu/T$ and $1/\mu=(\sum_i m_i^{-1})$ is the reduced mass of the system.
The contributions of non-standard annihilations 
have been discussed in {\it e.g}.~\cite{Aoki:2012ub}
for two and three component DM system with a $Z_2 \times Z'_2$ symmetry.

\section{Model A}
In the following, by extending the one-loop model in~\cite{Ma:2006km}
we study two of the two-loop radiative seesaw models with $Z_2 \times Z'_2$ symmetry.
We refer to them as ``model A" and ``model B". 
Owing to the $Z_2 \times Z'_2$ symmetry, there exist at least two extra stable electrically 
neutral particles.
The multicomponent DM system is realized in the model A, while one of the stable particles plays as
the dark radiation 
in the model B.

The matter content of the model A is shown in Table I.
In addition to the matter content of the SM model,
we introduce the right-handed neutrino $N_k$, 
an $SU(2)_L$ doublet scalar $\eta$, 
and two SM singlet scalars $\chi$ and $\phi$.
Note that the lepton number $L'$ of $N$ is zero.
The $Z_2 \times Z'_2 \times L'$ -invariant 
Yukawa sector and Majorana mass term for $N$ can be described by
\begin{equation}
\mathcal{L}_Y 
= Y^e_{ij} H^\dag L_i  l_{Rj}^c
 + Y^{\nu}_{ik}L_i \epsilon \eta N_{k} -\frac{1}{2} 
 M_{k} N_{k} N_{k} 
 + h.c. ~,
 \label{LY}
\end{equation}
where $i,j,k~(=1,2,3)$ stand for  the flavor indices.
The scalar potential $V$ is written as $V=V_{\lambda}+V_m$,
where 
\begin{align}
V_{\lambda}&=
\lambda_1 (H^\dag H)^2
 +\lambda_2 (\eta^\dag \eta)^2
 + \lambda_3 (H^\dag H)(\eta^\dag \eta)
+\lambda_4 (H^\dag \eta)(\eta^\dag H)
 \nonumber \\
& 
 + \gamma_1 \chi^4
 +\gamma_2 (H^\dag H)\chi^2
 + \gamma_3 (\eta^\dag \eta)\chi^2
 + \gamma_4 |\phi|^4
+ \gamma_5 (H^\dag H)|\phi|^2\nn\\
&+ \gamma_6 (\eta^\dag \eta)|\phi|^2
+ \gamma_7 \chi^2|\phi|^2
+ \frac{\kappa}{2} [\,(H^\dag \eta)\chi \phi+h.c.\,]~,
 \label{potential}\\
V_{m}&=m_1^2 H^\dag H + m_2^2 \eta^\dag \eta 
+ \frac{1}{2} m_3^2 \chi^2+ m_4^2 |\phi|^2
+ \frac{1}{2} m_5^2 [\, \phi^2+(\phi^*)^2\,]~.
\label{v2A}
\end{align}
The $Z_{2}\times  Z'_2$ is  the unbroken discrete symmetry
while the lepton number $L'$ is softly broken 
by the  last term in the potential $V_m$, the $\phi$ mass tem.
In the absence of this term, there will be no neutrino mass.
Note that  the ``$ \lambda_5$ term'', 
$(\lambda_5/2) (H^\dag \eta)^2$,  is also forbidden by $L'$.
A small $\lambda_5$ of the original model of Ma \cite{Ma:2006km}
is ``natural'' according to 't Hooft \cite{'tHooft:1979bh}, because the absence of 
$\lambda_5$ implies an enhancement of symmetry. 
In fact, if $\lambda_5$ is small at some scale, it remains
small for other scales as one can explicitly verify \cite{Bouchand:2012dx}.
Here we attempt to derive the smallness of $\lambda_5$
dynamically, such that the $\lambda_5$ term becomes calculable.

\begin{table}
\begin{center}
\begin{tabular}{|c|c|c|c|c|c|c|} 
\hline
field & statistics& $SU(2)_L$ & $U(1)_Y$ & $Z_2$ &$Z'_2$ & $L'$	\\ \hline
$L=(\nu_{L},l_L)$	& F& $2$ 	& $-1/2$ 	& $+$ & $+$ & $1$\\ \hline
$l^c_R$ 		& F	& $1$ 	& $1$		& $+$ & $+$ &$-1$	\\ \hline
$N$		& F		& $1$		& $0$ 	& $-$ & $+$ &$0$	\\ \hline
$H=(H^+,H^0)$ 	& B& $2$ 	& $1/2$ 	& $+$ & $+$	 &$0$\\ \hline
$\eta=(\eta^+,\eta^0) 	$ 	& B& $2$ 	& $1/2$ 	& $-$ & $+$ &$-1$	\\ \hline
$ \chi $     	& B  & $1$    & $0$        & $+$ & $-$	 &$0$\\ \hline
$ \phi $   	& B  & $1$    & $0$        & $-$ & $-$	 &$1$\\ \hline
\end{tabular}
\caption{\footnotesize{The matter content and the corresponding quantum numbers of the model A. 
}}
\end{center}
\label{contents}
\end{table}

\begin{figure}[t]
\includegraphics[width=8cm]{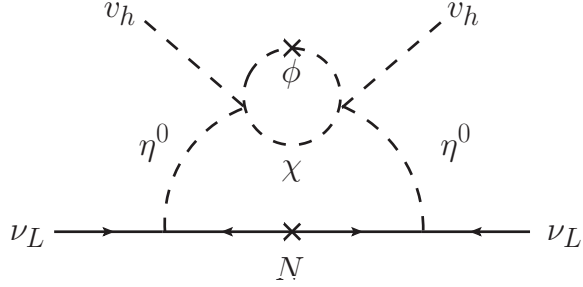}
\caption{\label{twoloop1}\footnotesize
Two-loop radiative neutrino mass of the model A.
}
\end{figure}

The Higgs doublet field $H$, the inert doublet field $\eta$ and the singlet scalar $\phi$ are
respectively parameterized as 
\begin{align}
H &= \left(\begin{array}{c}H^+ \\ (v_h+h+i G)/\sqrt{2}\\
\end{array}\right)~,~
\eta = \left(\begin{array}{c}\eta^+ 
\\ (\eta_{R}^0+i\eta_{I}^0)/\sqrt{2}\\
\end{array}\right)~,~\phi=(\phi_R+i \phi_I)/\sqrt{2}~,
\label{Higgs}
\end{align}
where $v_h$ is the vacuum expectation value.
The tree-level masses of the scalars are given by
\begin{align}
m_h^2 &=2 \lambda_1 v_h^2~, \\
m^2_{\eta^\pm}&=m_2^2+\lambda_3 v_h^2/2~, \\
m^2_{\eta_R^0}&=m^2_{\eta_I^0}=m_2^2+(\lambda_3+\lambda_4) v_h^2/2~, \label{eta_mass} \\
m^2_{\phi_R}&= m_4^2+m_5^2+\gamma_5 v_h^2, \\
m^2_{\phi_I}&=m_4^2-m_5^2+\gamma_5 v_h^2~,\\
m^2_{\chi}&=m_3^2+\gamma_2 v_h^2~.
\end{align}
Althogh
the tree-level mass of $\eta^0_R$ is the same as that of $\eta^0_I$ as shown in (\ref{eta_mass}),
the degeneracy is lifted at the one-loop level via the effective $\lambda_5$ term:
\begin{align}
\lambda_5^{\rm eff} 
&\sim
-\frac{\kappa^2}{128\pi^2}
\frac{m_5^2}{m_{\phi_R}^2-m_{\chi}^2}
\left[
1-
\frac{m^2_{\chi}}{m^2_{\phi_R}-m_\chi^2}
\ln \frac{m^2_{\phi_R}}{m_\chi^2}
\right]~\mbox {for}~ m_5\ll m_{\phi_R}~.\label{l5}
\end{align}
This correction is embedded into 
 the two-loop diagram to generate the neutrino mass (see Fig.~\ref{twoloop1}).
The 3$\times$3 neutrino mass matrix ${\cal M}_\nu$ can be given by
\be
({\cal M}_\nu)_{ij} 
&=&
\left(\frac{1}{16\pi^2}\right)^2
\frac{\kappa^2 v_h^2 }{16}
 \sum_k Y^\nu_{ik}Y^\nu_{jk}M_k ~
 \int_0^\infty dx ~\frac{x}{(x+m_{\eta^0}^2)^2(x+M_k^2)}\nn\\
& &\times \int_0^1 dz \ln\left[
 \frac{z m_\chi^2+(1-z)m_{\phi_I}^2+z(1-z) x}{z m_\chi^2+(1-z)m_{\phi_R}^2+z(1-z) x}
 \right]~,
\label{mnu0}
\ee
where we have assumed that 
$m_{\eta^0}=m_{\eta_R^0}\simeq m_{\eta_I^0}$.

Using $\lambda_5^{\rm eff}$ given in (\ref{l5}), 
the neutrino mass matrix can be approximated as
\begin{align}
({\cal M}_\nu)_{ij} 
&\sim
\frac{\lambda_5^{\rm eff} v_h^2}{32 \pi^2}  
\sum_k 
\frac{Y^\nu_{ik} Y^\nu_{jk} M_k}{m_{\eta^{0}}^2 - M_k^2}
\left[ 1 - \frac{M_k^2}{m_{\eta^{0}}^2 - M_k^2}
 \ln \frac{m_{\eta^{0}}^2}{M_k^2} \right].
\label{mnu}
\end{align}
We see from  (\ref{l5}) and (\ref{mnu}) that
the neutrino mass matrix ${\cal M}_\nu$ is proportional to 
$|Y^\nu \kappa|^2 m_5^2$.
When $m_{\chi}$, $m_{\phi_R}$, $m_{\eta^0}$, $M_k
 \sim \mathcal{O}(10^2) ~\mathrm{GeV}$,
 for instance, implies that 
 $|Y^\nu \kappa| m_5 \sim \mathcal{O}(10^{-1}) ~\mathrm{GeV}$
  to obtain the neutrino mass scale of $ \mathcal{O}(0.1) ~\mathrm{eV}$.
  With the same set of the parameter values we find that $\lambda^{\rm eff}_5
  \sim 10^{-6}$, where the smallness 
  $\lambda^{\rm eff}_5$ is a consequence of the radiative generation
  of this coupling.
  As we will see, the product $ |Y^\nu \kappa|$ enters into the semi-annihilation
  of DM particles which produces monochromatic neutrinos,
  while the upper bound of $|Y^\nu|$ follows from the $\mu \to e \gamma $ constraint.

\subsection{Multicomponent dark matter system}

In the model A there are three type of dark matter candidates ;
$N_{1}$ (the lightest among $N_{k}$'s) or $\eta^0_R$ (or $\eta^0_I$) with $(Z_2, Z'_2)=(-,+)$, 
$\chi$ with $(Z_2, Z'_2)=(+,-)$ and $\phi_R$ (or  $\phi_I$) with $(Z_2, Z'_2)=(-,-)$. 
For $(Z_2, Z'_2)=(-,+)$ there are two candidates. 
In the following discussions we assume that $N_1$ is a DM  candidate~\cite{Aoki:2014lha}.
The other possibility, $\eta_R^0$-DM, is discussed in \cite{Aoki:2013gzs}.

\begin{figure}[t]
 \includegraphics[width=13cm]{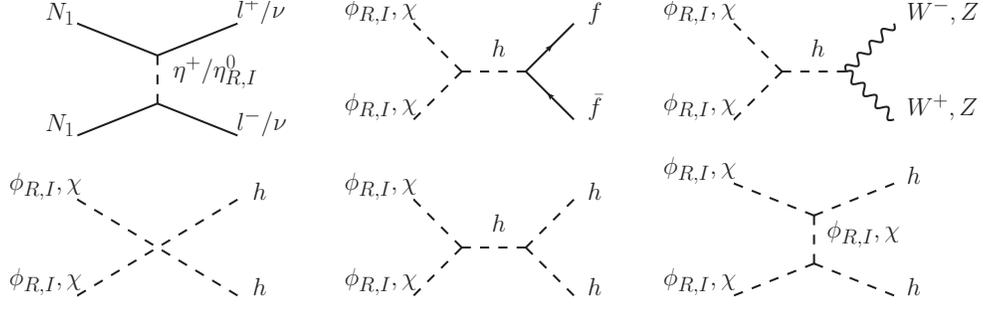}
	\caption{The diagrams for the standard annihilation processes. }
	\label{fig:stand}
\end{figure}

\begin{figure}[h]
	\includegraphics[width=13cm]{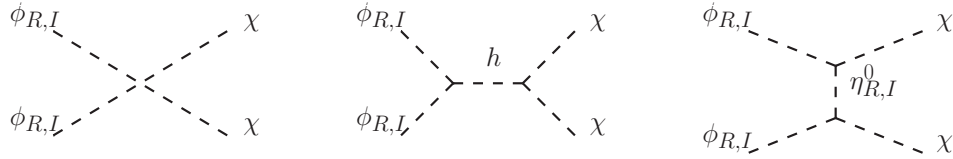}
	\caption{The diagrams for the DM conversion processes.}
	\label{fig:DM_con}
\end{figure}

\begin{figure}[h]
	\includegraphics[width=13cm]{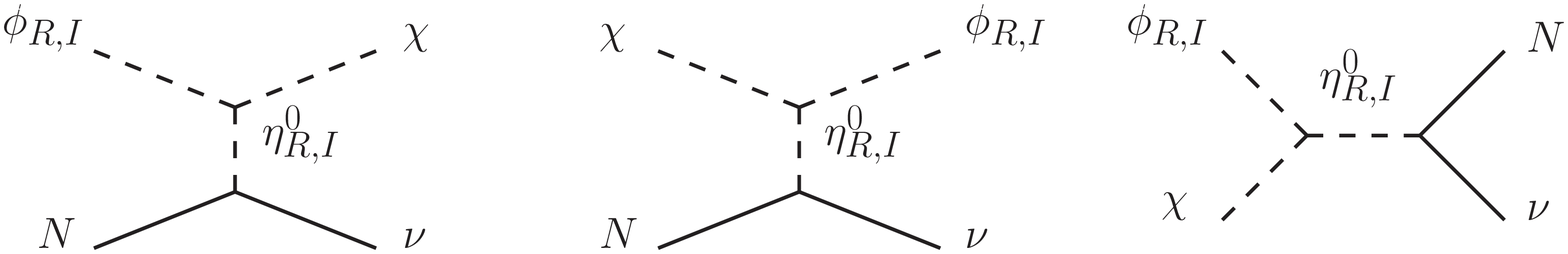}
	\caption{The diagrams for semi-annihilation process.}
	\label{fig:semi}
\end{figure}

We discuss the three DM system of  $N_1 ,~ \phi_R,~ \chi$.
There are three types of DM annihilation process:
\begin{align}
\mathrm{Standard~annihilation:}&~
N_1N_1 \rightarrow XX',~~\phi_R \phi_R \rightarrow XX',~~\chi\chi \rightarrow XX',~\\
\mathrm{DM~conversion:}&~\phi_R \phi_R \rightarrow \chi\chi,~\\
\mathrm{Semi}\rm{\mathchar`-annihilation:}&~
N_1 \phi_R \rightarrow \chi \nu,~~
\chi N_1  \rightarrow \phi_R \nu,~~
\phi_R \chi \rightarrow N_1 \nu.
\end{align}
Here we have assumed $m_{\phi_R} > m_\chi$ and $m_{\phi_R} + m_\chi < M_{2,3}$.
Moreover, since the mass difference between $\phi_R$ and $\phi_I$ is controlled by the lepton-number breaking mass $m_5$, which is assumed to be much smaller than $m_{\phi_R}$.
Then $m_{\phi_R}$ and $m_{\phi_I}$ are practically degenerate and the contribution of $\phi_I$ to
the annihilation processes during the decoupling of DMs is nonnegligible. 
The diagrams for annihilation processes which enter the Boltzmann equation
are shown 
in Figs.~\ref{fig:stand}-\ref{fig:semi}.
Since the reaction rate of the conversion between $\phi_R$ and $\phi_I$ can reach chemical equilibrium
during the decoupling of DMs, 
we can sum up the number densities of $\phi_R$ and $\phi_I$
and compute the relic abundance of $\Omega_{\phi_R}h^2$~\cite{Aoki:2014lha}.
%

In the multicomponent DM scenario, 
the effective cross section off the nucleon is given by
\begin{eqnarray}
\sigma^{\mathrm{eff}}_i
=
\sigma_i 
\left( \frac{ \Omega_i h^2 }{ \Omega_{\mathrm{total}} h^2 } \right)
~. \label{ddeff}
\end{eqnarray}
In our model, only $\chi$ and $\phi_R$ scatter with the nucleus, and  
the right-handed neutrino $N_1$ does not interact with nucleus at tree level. 
So we can  neglect the  $N_1$ contribution at the lowest order
in perturbation theory.  The effective cross sections of $\phi_R$ and $\chi$ are expressed as
\begin{align}
\sigma^{\mathrm{eff}}_{\chi}
=
\sigma_{\chi}
\left( \frac{ \Omega_{\chi} h^2 }{ \Omega_{\mathrm{total}} h^2 } \right) \,,
~~~
\sigma^{\mathrm{eff}}_{\phi_R}
=
\sigma_{\phi_R}
\left( \frac{ \Omega_{\phi_R} h^2 }{ \Omega_{\mathrm{total}} h^2 } \right) \,,
 \label{ddsigma}
\end{align}
where $\sigma_{\chi}$ and $\sigma_{\phi_R}$ are the spin independent cross sections and given by 
\begin{align}
\sigma_{\chi}
&=
\frac{1}{\pi}
\left(  \frac{ \gamma_2 \hat{f} m_N }{ m_{\chi} m_h^2  }  \right)^2
\left(  \frac{ m_N m_{\chi} }{ m_N + m_{\chi} }  \right)^2 ~,
\\
\sigma_{\phi_R}
&=
\frac{1}{\pi}
\left(  
\frac{ (\gamma_5/2) \hat{f} m_N }{ m_{\phi_R} m_h^2 }  
\right)^2
\left(  \frac{ m_N m_{\phi_R} }{ m_N + m_{\phi_R} }  \right)^2 ~. 
\end{align}
Here $\hat{f} \sim 0.3$ is the usual nucleonic matrix element \cite{Ellis:2000ds} and $m_N$ is nucleon mass.

The upper bounds on the cross section off the nucleon is obtained by 
LUX \cite{Akerib:2016vxi} and XENON1T \cite{Aprile:2017iyp}. 
In the cases of one-component DM system
of a real or complex scalar boson,  
those experimental results give the strong constraint on the masses of those DM particles;
the allowed DM mass region is $\simeq m_h/2$ and $\gsim 1$ TeV~\cite{Wu:2016mbe}.
In the model A with the multicomponent DM system, the constrains on the cross sections off the nucleon for $\chi$ and $\phi_R$
are also relatively severe.
As a benchmark we take the mass of $\chi$ as $m_\chi=m_h/2$ while vary the mass of $\phi_R$ in the following analysis \footnote{Two singlet scalar DM scenario in $Z_2 \times Z_2'$ model has been explored in detail in Ref.~\cite{Bhattacharya:2016ysw}. }.

In the original one-loop neutrino mass model in  \cite{Ma:2006km}, the relic density of $N_1$ tends to be larger than the observational value \cite{{Kubo:2006yx}}. 
The additional contributions coming from the semi-annihilation can enhance the annihilation rate for $N_1$ so that the $N_1$ DM contribution to the total relic abundance can be suppressed. 
This situation is realized for $ M_{1} > m_{\phi_R}, m_{\chi}$ 
as can be seen later.

As the benchmark set we take the following values for the parameters.
\begin{align}
&m_{\chi}=62 ~\mathrm{GeV},   ~~M_1= 300 ~\mathrm{GeV}, \label{BM1}\\
&m_{\eta^0_R}=m_{\eta^0_I}=m_{\eta^+}=  m_{\chi} +m_{\phi_R} - 10~ \mathrm{GeV}, \label{mdiff}\\
&m_{\phi_I}= m_{\phi_R}+5~\mathrm{GeV}\,, \\
&\gamma_2=0.004,  ~~\gamma_5=0.05,  ~~\gamma_7=0.17\,, \\ 
&  \kappa=0.4, 	~~ Y_{ij}^\nu =0.01 \,. 
\label{BM2}
\end{align}
The masses of heavier right-handed neutrinos are $M_{2}=M_3=1$ TeV.
The mass differences between $m_{\eta^0_{R}}$ and the sum of $m_{\chi}$ and $m_{\phi_R}$ are so chosen that no resonance appears in the $s$-channel of the semi-annihilation in Fig.~\ref{fig:semi} (right).
The benchmark set satisfies the constraints from the perturbativeness,
the stability conditions of the scalar potential~\cite{Aoki:2013gzs,Aoki:2014lha}, 
the lepton flavor violation (LFV) such as $\mu\to e\gamma$~\cite{TheMEG:2016wtm}
and the electroweak precision measurements \cite{Barbieri:2006dq,Baak:2014ora}.
It is noted that
$\kappa$ is bounded as $|\kappa |\lesssim 0.4$ by the perturbativeness and
the stability conditions~\cite{Aoki:2013gzs,Aoki:2014lha}.

\begin{figure}[t]
\centering
		\includegraphics[width=10cm]{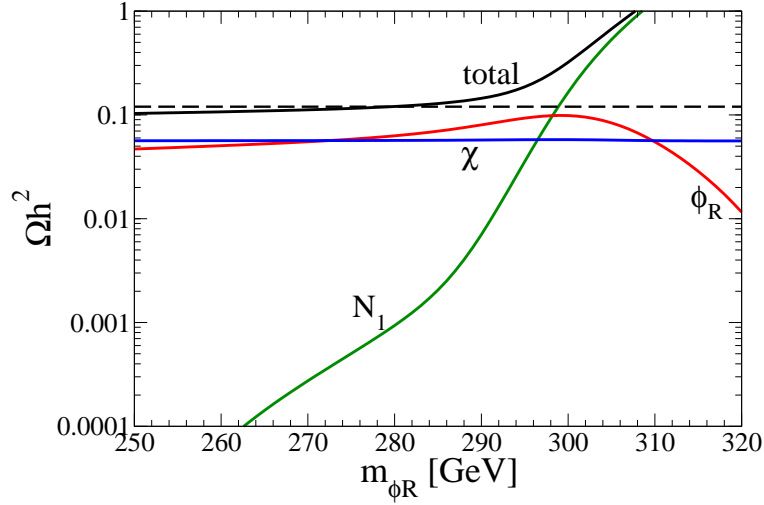}
	\caption{Relic abundances $\Omega_{\chi}h^2$, $\Omega_{\phi_R}h^2$ and $\Omega_{N_1}h^2$ and the total relic
	abundance $\Omega_{\rm total} h^2$
 as a function of  $m_{\phi_R}$. The relevant masses and couplings are taken as in Eqs.(\ref{BM1})-(\ref{BM2}).
 The horizontal dashed line stands for the observed value $\Omega_{\rm obs} h^2 \sim 0.12$.}
\label{fig:relic}
\end{figure}
%
%

\begin{figure}
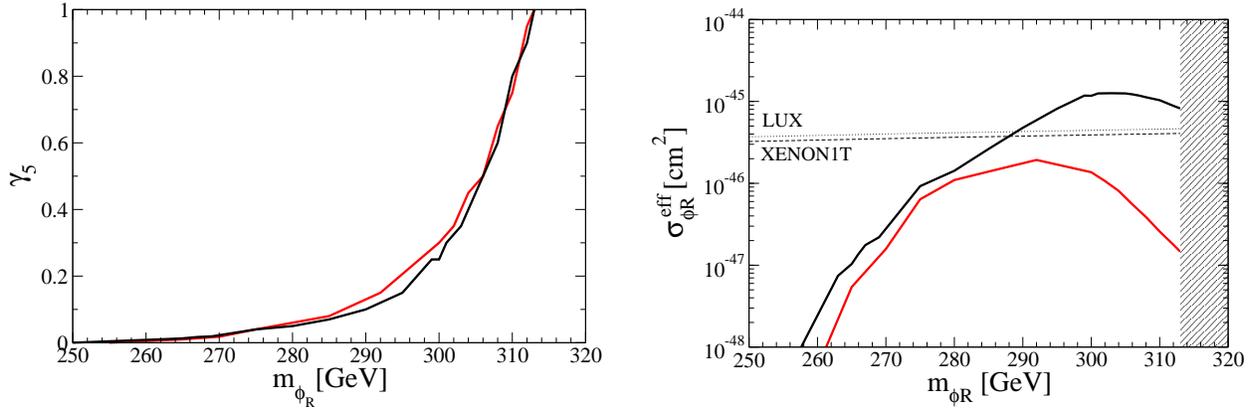

\centering
\vspace{9mm}
\includegraphics[width=7.9cm]{Figs/modelA_gamma.eps}
~~~~\includegraphics[width=7.9cm]{Figs/modelA_DD.eps}
\caption{
\footnotesize 
Left 
: Contour plot for the total relic density $\Omega_{\rm total} h^2 \sim 0.12$. 
Right
:The relation between the $m_{\phi}$  and 
the effective cross sections given in Eq.~(\ref{ddsigma}).
 The black dot and dashed lines show the upper limit of the spin independent cross section 
 off the nucleon given by 
 LUX~\cite{Akerib:2016vxi} and XENON1T \cite{Aprile:2017iyp}, respectively.
 The hatched region is excluded by perturbativity.
 In both panels, we take two values, 10 GeV (black line)
and 1 GeV (red line), 
for the mass difference between $m_{\eta^0_R}$ and $m_\chi +m_{\phi_R}$. }
 \label{fig:modelA_cont} 
\end{figure}
%

\begin{figure}
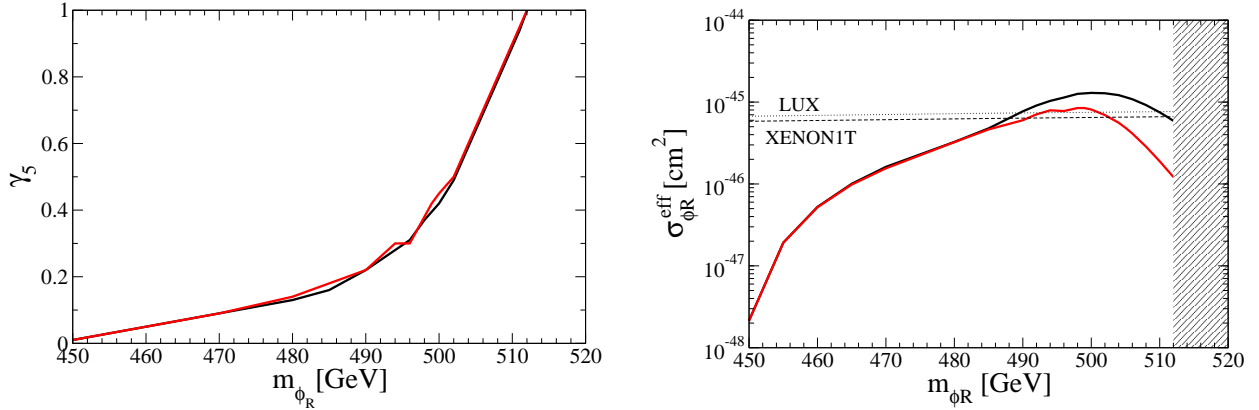

\centering
\vspace{8mm}
\includegraphics[width=7.9cm]{Figs/modelA_gamma_500.eps}
~~~~\includegraphics[width=7.9cm]{Figs/modelA_DD_500.eps}
\caption{
\footnotesize 
The same as Fig.~\ref{fig:modelA_cont} but for  $M_1=500$ GeV and $\gamma_7=0.28$. 
 }
 \label{fig:modelA_cont_500} 
\end{figure}
%

Figure~\ref{fig:relic} shows the relic abundances of $\Omega_{\chi}h^2$, $\Omega_{\phi_R}h^2$ and $\Omega_{N_1}h^2$ and the  total relic abundance $\Omega_{\rm total} h^2(=\Omega_{\chi}h^2+\Omega_{\phi_R}h^2+\Omega_{N_1}h^2)$
as a function of  $m_{\phi_R}$ for the benchmark set. The horizontal dashed line stands for the observed value $\Omega_{\rm obs} h^2 \sim 0.12$.
It is shown that 
the relic abundance of the $\chi$ is $\Omega_\chi \simeq \Omega_{\rm obs}/2$.
When $\phi_R$ is lighter than $N_1$, the semi-annihilation tends to decrease the relic abundance of $N_1$.
For the benchmark set, the total relic abundance is consistent with the observed value around $m_{\phi_R} \simeq 280$ GeV.

The left panel in Fig.~\ref{fig:modelA_cont} shows the contour plot for the $m_{\phi_R}$-$\gamma_5$ plane where  
the total relic density of DM can be made consistent with the observed value $\Omega_{\rm obs} h^2 \sim 0.12$.
We take two values, 10 GeV (black line)
and 1 GeV (red line), 
for the mass difference between $m_{\eta^0_R}$ and $m_\chi +m_{\phi_R}$ in~(\ref{mdiff}). 
The other parameters are taken as the same in Eqs.(\ref{BM1})-(\ref{BM2}).
We can see the scalar coupling $\gamma_5$ increases drastically 
as $m_{\phi_R}$ increases for  $m_{\phi_R} \gsim 290 ~\mathrm{GeV}$.
It is because the relic density of the $N_1$ DM, $\Omega_{N_1} h^2$, becomes significant 
 for $m_{\phi_R} \gsim 290$ GeV,  so that $\Omega_{\phi_R} h^2$ should be drastically suppressed.
The scalar couplings of DM particles with the SM Higgs boson, $\gamma_2$ and $\gamma_5$, and the DM masses are constrained by the DM direct detection experiments.
For the $\chi$ DM, the effective cross section off nucleon $\sigma^{\mathrm{eff}}_{\chi}$ in Eq.~(\ref{ddsigma})
is $\sigma^{\mathrm{eff}}_{\chi}\sim10^{-47}$ cm$^2$ for the benchmark set.
It is an order of magnitude smaller than the current experimental bound.
For the $\phi_R$ DM,
the right panel in Fig.~\ref{fig:modelA_cont} shows the relation between $m_{\phi_R}$ and 
the effective cross section $\sigma^{\mathrm{eff}}_{\phi_R}$
for $(m_\chi + m_{\phi_R})-m_{\eta^0_R}$=10 GeV (black line) and 1 GeV (red line),
where the DM relic abundance is consistent with  the observation. 
The plot corresponds to the parameter space in the left panel in Fig.~\ref{fig:modelA_cont}. 
The dot and dashed lines indicate the upper bounds of LUX and XENON1T, respectively.
The hatched region is excluded by perturbativity.
Although the scalar coupling $\gamma_5$ becomes large for $m_{\phi_R} \gsim 290 ~\mathrm{GeV}$
and then
the cross sections off the nucleon $\sigma_{\phi_R}$ becomes large, 
the effective cross section $\sigma_{\phi_R}^{\rm eff}$ decreases
for $m_{\phi_R} \gsim M_{1} (= 300$ GeV), 
since the abundance of $\phi_R$ decreases.
For the case of $(m_\chi + m_{\phi_R})-m_{\eta^0_R}$=10 GeV,
it can be seen that the mass region $ 288 ~\mathrm{GeV} \lsim m_{\phi_R}$ is excluded by LUX and XENON1T data.
On the other hand, there are no constraints from the direct DM search experiments on the mass of $\phi_R$
for the case of  $(m_\chi + m_{\phi_R})-m_{\eta^0_R}$=1 GeV.
This is because
the relic abundance of $\phi_R$ becomes much smaller 
by the large contribution 
from the $s$-channel process
of the semi-annihilation.
Figure~\ref{fig:modelA_cont_500}  shows the same as in Fig.~\ref{fig:modelA_cont} but  
for  $M_1=500$ GeV and $\gamma_7=0.28$. 
From the right panel in Fig.~\ref{fig:modelA_cont_500},
we see that 485 (490) GeV $\lsim m_{\phi_R} \lsim$ 510 (502) GeV is excluded by the direct detection experiments
for $(m_\chi + m_{\phi_R})-m_{\eta^0_R}$=10 (1) GeV.

Before we go to discuss indirect detection, we summarize 
the  parameter space, in which  
a correct value of the total relic DM abundance
$\Omega_{\rm total} h^2$ can be obtained without contradicting 
the constraint from the direct detection experiments.
As in the case of the single SM singlet DM, the constraint
is in fact very severe: The mass of $\chi$ has to be very close to
$m_h/2$, and $\gamma_2$ (the Higgs portal coupling) also has to be close to $0.004$ for an adequate amount of 
$\Omega_\chi$.
However, as for $m_{\phi_R}$ and  $\gamma_5$, 
there exist a certain allowed region.
The allowed region in the $m_{\phi_R}$-$\gamma_5$ plane 
is controlled by 
the semi-annihilation (especially, the last diagram in  Fig.~\ref{fig:semi}, which is sensitive
to the mass relation (28)) and the DM conversion (especially 
the right diagram in Fig.~\ref{fig:DM_con}, which is sensitive to $\gamma_7$).
If we increase the mass of the 
right-handed neutrino DM, the mass of $\phi_R$ increases, but how the allowed range in the 
$m_{\phi_R}$-$\gamma_5$ plane emerges remains the same. 
If we take the larger $\gamma_7$, {\it e.g.} $\gamma_7=0.28$, in Fig.~\ref{fig:modelA_cont},
the allowed region for $m_{\phi_R}$
becomes narrower as $295~\mathrm{GeV} \lsim m_{\phi_R} \lsim 300~\mathrm{GeV}$.
The smaller $m_{\phi_R}$ ($\lsim 295~\mathrm{GeV}$) is excluded by $\Omega_{\rm total} < \Omega_{\rm obs}$ due to the larger DM conversion {\it i.e.} the larger annihilation process of $\phi_R\phi_R\to \chi\chi \to XX$.

\subsection{Indirect detection} 
For indirect detections of DM 
 the SM particles produced by the annihilation of DM are searched.
Because the semi-annihilation produces a SM particle, this process can 
serve for an indirect detection. 
In our model, especially, the SM particle from the semi-annihilation process as
shown in Fig.~\ref{fig:semi} is neutrino which has a monochromatic energy spectrum.
Therefore,  we consider below the neutrino flux from the Sun 
\cite{Silk:1985ax,
Griest:1986yu, 
Ritz:1987mh, 
Kamionkowski:1991nj,
Jungman:1995df} 
as a possibility to detect the semi-annihilation process of DMs.

The DM particles are captured in the Sun losing their 
kinematic energy through scattering with the nucleus. 
Then captured DM particles annihilate each other.
The time dependence of the  number of DM $n_i$ in the Sun is given by
\cite{Kamionkowski:1991nj, Ritz:1987mh,Griest:1986yu,Jungman:1995df}
\begin{align}
\dot{n}_i
&=
C_i
 - C_A(ii \rightarrow \mathrm{SM})n_i^2
 - \sum_{m_i>m_j}C_A(ii \rightarrow jj)n_i^2
 - C_A(ij \rightarrow k \nu)n_i n_j~,
 \label{Boltzsun}
\end{align}
where $i, j ,k =\chi, \phi_R, N_1$ and
$C_i$ is the capture rates in the Sun:
\begin{align}
&C_\chi \simeq 
 2.5\times 10^{18} \mbox{s}^{-1} f(m_{\chi})\left( \frac{\hat{f}}{0.3} \right)^2 
\left(\frac{\gamma_{2}}{0.004} \right)^2 
\left(\frac{60~\mathrm{GeV}}{m_{\chi}} \right)^2 
\left(\frac{125~\mathrm{GeV}}{m_{h}} \right)^4 
\left(\frac{\Omega_{\chi} h^2}{ \Omega_{ \mathrm{total} }{h^2}} \right)
,\\
&C_{\phi_R} \simeq 
 6.2\times 10^{17} \mbox{s}^{-1} f(m_{\phi_R})\left( \frac{\hat{f}}{0.3} \right)^2 
\left( \frac{\gamma_{5}}{0.02} \right)^2 
\left( \frac{300~\mathrm{GeV}}{m_{\phi_R}} \right)^2 
\left( \frac{125~\mathrm{GeV}}{m_{h}} \right)^4 
\left( \frac{\Omega_{\phi_R} h^2}{ \Omega_{ \mathrm{total} }{h^2}}   \right)
,\\
&C_{N_1} =  0~,
\end{align}
and $C_A$'s are the annihilation rate:
\begin{align}
C_A(ij \rightarrow \bullet)&=
\frac{ \langle \sigma(ij \rightarrow \bullet)v \rangle }
{ V_{ij} }
~,~~~
V_{ij} = 5.7 \times 10^{27}
\left( \frac{100~\mathrm{GeV}}{\mu_{ij}} \right)^{3/2} \mathrm{cm}^3 ~.
\end{align}
Here $f(m_{i})$ depends on the form factor of the nucleus, elemental abundance, kinematic suppression of the capture rate, etc.,
varying ${\mathcal O}(0.01 - 1)$ depending on the DM mass \cite{Kamionkowski:1991nj,Jungman:1995df}.
$V_{ij}$ is an effective volume of the Sun with $\mu_{ij}=2m_i m_j / (m_i+m_j)$
 in the non-relativistic limit. 
In the Eq.(\ref{Boltzsun}) we have neglected the DM production processes such as $jj \rightarrow ii$ and $jk \rightarrow iX$ because the kinetic energy of the produced particle $i$ is much larger than  that corresponding to
the escape velocity from  the Sun, i.e. $\sim 10^3$ km/s
\cite{Griest:1986yu,Agrawal:2008xz}. 
Consequently,  the number of the right-hand neutrino DM cannot increase in the Sun, 
and hence the semi-annihilation process, $\phi_R\chi \rightarrow N_1 \nu$, is the only neutrino production process
\footnote{There are also neutrinos having continuous energy spectrum from the decay of SM particles
produced by the standard annihilation. The upper bounds for the production rates of the SM particles are given in~\cite{Agrawal:2008xz, Aartsen:2016zhm}.},
where
its reaction rate as a function of $t$ is given by 
$\Gamma(\phi_R \chi \rightarrow N \nu;t) = C_A(\phi_R \chi \rightarrow N_1 \nu)n_{\phi_R}(t) n_\chi (t) $.
\begin{figure}
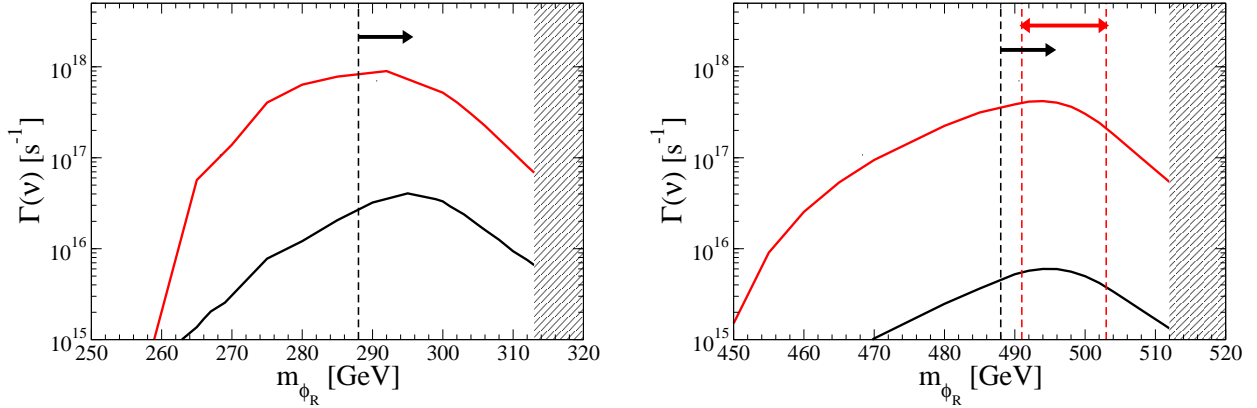

\begin{center}
\includegraphics[width=7.9cm]{Figs/modelA_ID.eps}
~~~~\includegraphics[width=7.9cm]{Figs/modelA_ID_500.eps}
	\caption{ 
	\footnotesize 
	The neutrino production rate $\Gamma(\nu) =\Gamma(\phi_R \chi \rightarrow N \nu;t_0)$
	in the Sun 
	against the $\phi_R$ DM mass for $M_1=300$ GeV (left) and 500 GeV (right). 
	The parameter space, as well as the meaning of colors of the lines
	in the left and right panel, are the same as in Fig.~\ref{fig:modelA_cont} and \ref{fig:modelA_cont_500}, respectively.
	The hatched region is excluded by perturbativity.
Arrows indicate the excluded regions by the direct detection experiments.
	}
\label{nufluxreson}
\end{center}	
\end{figure}

Figure~\ref{nufluxreson} shows the $m_{\phi_R}$ dependence 
of the neutrino production rate today $\Gamma(\phi_R \chi \rightarrow N \nu; t_0)$, 
where $t_0=1.45\times 10^{17}$s is the age of the Sun, for the same parameter space
as in Fig.~\ref{fig:modelA_cont}
(Fig.~\ref{nufluxreson} (left)) and in Fig.~\ref{fig:modelA_cont_500} (Fig.~\ref{nufluxreson} (right)).
The hatched region is excluded by perturbativity.
Arrows indicate the excluded regions by the direct detection experiments.
For $m_{\phi_R} \gsim M_{1}$ 
where the relic abundance of $N_1$ dominates that of $\phi_R$,
the neutrino production rate decreases since the capture rate of the $\phi_R$ becomes small.
As we can see from Fig.~\ref{fig:semi} a resonance effect for
the $s$-channel annihilation process can be achieved if 
$m_{\eta^0_R} \simeq m_{\phi_R}+m_\chi$. 
Then the smaller neutrino mass difference $m_{\eta^0_R} -(m_{\phi_R}+m_\chi )$
gives the larger neutrino production rate.
For the case of $m_{\eta^0_R} -(m_{\phi_R}+m_\chi )$= 1 GeV, the rate 
$\Gamma(\phi_R \chi \rightarrow N \nu; t_0)$ reaches about 
$10^{18}$ s$^{-1}$ at $m_{\phi_R}\simeq 290$ GeV for $M_1=300$ GeV 
and $4\times 10^{17}$ s$^{-1}$
at $m_{\phi_R}\simeq 490$ GeV for $M_1=500$ GeV, respectively. 

The upper limits on the DM DM $\to XX'$ from the Sun are given 
 by IceCube experiment \cite{Aartsen:2016zhm}. 
For instance, 
the upper limit on the annihilation rate of the DM of 250 (500) GeV
into $W^+W^-$  is $1.13\times 10^{21}$ $(2.04\times 10^{20})$ s$^{-1}$ and that into $\tau^+\tau^-$ is $5.99\times 10^{20}$ $(7.96\times 10^{19})$ s$^{-1}$, 
which is at least $10^{2}$ times larger than the rate $\Gamma (\nu)$ shown in 
Fig.~\ref{nufluxreson}.
Note however that the energy spectrum of the neutrino flux produced by the $W$ or $\tau$ decay is different from the monochromatic neutrino. With  an increasing resolution of energy and angle
 the  chance for the observation of the semi-annihilation
 and hence of a multicomponent nature of DM can increase.


\section{Model B}
Neutrinos have always played consequential roles in cosmology
(see \cite{Lee:1977ua}, and also \cite{Kolb:1990vq}
and references therein).
While they play a role as hot dark matter,
the mechanism of their mass
generation is directly connected to cosmological problems 
such as baryon asymmetry of the Universe \cite{Fukugita:1986hr}
and dark matter~\cite{Krauss:2002px,Ma:2006km,Nasri:2001ax,
Ma:2007yx, Aoki:2013gzs, Aoki:2014lha,
Ma:2006uv, Cai:2017jrq,Kubo:2006yx}.
Resent cosmological 
observations with increasing accuracy 
\cite{Rozo:2009jj,Riess:2011yx,Heymans:2013fya,Ade:2013zuv}
provide useful hints
 on how to extend  the neutrino sector.
Here we propose an extension of the 
neutrino sector such that 
 the tensions
among resent different cosmological observations
can be alleviated.
The tensions have emerged since the first  Planck result
 \cite{Ade:2013zuv}
 in the Hubble constant $H_0$ and in  
 the density variance $\sigma_8$ in spheres of radius $8h^{-1}$ Mpc:
 The Planck values of $1/H_0$  and 
$\sigma_8$ are slightly larger
than those obtained from the observations of the local Universe such as Cepheid variables \cite{Riess:2011yx} and 
the Canada-France- Hawaii Telescope Lensing Survey
\cite{Benjamin:2012qp}, respectively.
The Planck galaxy cluster counts \cite{Ade:2013lmv} 
and also the Sloan Digital Sky Survey  data \cite{Rozo:2009jj}
yield a smaller  $\sigma_8$.

It has been recently suggested
\cite{Ade:2013lmv,Hamann:2013iba,Feng:2017nss} that these tensions can be
alleviated if the number $N_{\rm eff }$ of the 
 relativistic species  in the young Universe is slightly larger than 
 the standard value $3.046$ and  the mass of the extra relativistic
specie is  of ${\cal O}(0.1)$ eV \cite{Feng:2017nss}.
Here we suggest a radiative  generation
mechanism of the neutrino mass, which is directly connected to 
the existence of a stable DM particle and also  a non-zero
$\Delta N_{\rm eff }=N_{\rm eff }-3.046$.

The matter content of the model is shown in Table II.
It is a slight modification of the model A:
$\chi$ in this model is a Majorana fermion.
The $Z_2 \times Z'_2 \times L'$ -invariant 
Yukawa  sector (the quark sector is suppressed) is described by
the Lagrangian 
\begin{align}
\mathcal{L}_Y 
&= Y^e_{ij} H^\dag L_i  l_{Rj}^c
 + Y^{\nu}_{ij}L_i \epsilon \eta N_{j}
+  Y^\chi_{ij} \, N_{i} \chi_j \phi -
  \frac{1}{2}M_{\chi_k}\chi_{k} \chi_{k}
 + h.c. ~,
 \label{LY}
\end{align}
 where $i,j,k=1,2,3$, and we have assumed
 without loss of generality that the $\chi$ mass term is diagonal.
 We also assume that $Y^e_{ij}$ have only diagonal elements.
The most general renormalizable form of the $Z_2 \times Z'_2\times L'$-invariant  scalar potential  is given by
\begin{align}
V_{\lambda}
&=
\lambda_1 (H^\dag H)^2
 +\lambda_2 (\eta^\dag \eta)^2
 + \lambda_3 (H^\dag H)(\eta^\dag \eta)
+\lambda_4 (H^\dag \eta)(\eta^\dag H)
+ \frac{\lambda_5}{2}[\, (H^\dag \eta)^2+h.c.\,]\nn\\
  & +
  \gamma_2 (H^\dag H)|\phi|^2
 + \gamma_3 (\eta^\dag \eta)|\phi|^2
+\gamma_4 |\phi|^4,
  \label{potential}
\end{align}
and the mass terms are
\begin{align}
V_{m}&=m_1^2 H^\dag H + m_2^2 \eta^\dag \eta 
+ m_3^2 |\phi|^2
-  \frac{m_4^2}{2} [\, \phi^2+(\phi^*)^2\,]~,
\label{vm}
\end{align}
where the $m_4$  term in (\ref{vm}) breaks $L'$ softly.
\begin{table}
\begin{center}
\footnotesize{
\begin{tabular}{|c|c|c|c|c|c|c|} 
\hline
field & statistics& $SU(2)_L$ & $U(1)_Y$  
 & $Z_2$ &$Z'_2$& $L'$
\\ \hline
$L=(\nu_L, l_L)$	& F& $2$ 	& $-1/2$  & $+$ & $+$&$1$\\ \hline
$l^c_R$ 		& F	& $1$ 	& $1$  &$+$ & $+$& $-1$  \\ \hline
$N$		& F		& $1$& $0$&$-$ & $+$ & $-1 $ \\ \hline
$H=(H^+,H^0)$ 	& B& $2$ 	& $1/2$ & $+$ & $+$ & $0$  \\ \hline
$\eta=(\eta^+,\eta^0) 	$ 	& B& $2$ 	& $1/2$ & $-$ 
& $+$&$0$\\ \hline
$ \chi $     	& F  & $1$    & $0$ &$+$ & $-$ & $0$  \\ \hline
$ \phi $   	& B  & $1$    & $0$ &$-$ & $-$ &$1$ \\ \hline
\end{tabular}
\label{tab2}}
\caption{\footnotesize{The matter content of the model B and
the corresponding quantum numbers. }}
\end{center}
\end{table}
The scalar fields 
$H$, $\eta$ and $\phi$
are defined in (\ref{Higgs}).
Since we assume that the discrete symmetry $Z_2 \times Z'_2$  is unbroken, the scalar fields above do not mix with other,
so that their tree-level masses can be simply expressed:
\begin{align}\label{massceta}
~m^2_{\eta^\pm}&=m_2^2+\lambda_3 v_h^2/2~,\\
m^2_{\eta_R^0}&=m_2^2+\left(\lambda_3+\lambda_4
+\lambda_5\right) v_h^2/2~, \\
m^2_{\eta_I^0}&=m_2^2+\left(\lambda_3+\lambda_4
-\lambda_5\right) v_h^2/2~, \\
m^2_{\phi_R} &= m_3^2-m_4^2+\gamma_2 v_h^2/2~, \\
m^2_{\phi_I}&=m_3^2+m_4^2+\gamma_2 v_h^2/2~.
\label{masses}
\end{align}

\begin{center}
\begin{figure}
    \includegraphics[width=9cm]{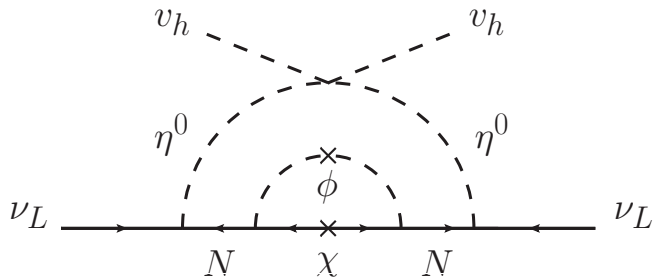}
\caption{\label{twoloop2}\footnotesize
Two-loop radiative neutrino mass of the model B.
The upper cross means the soft breaking mass term $m_4^2$, which 
should indicate that there are $\phi_R$ and $\phi_I$ loops in the inner one-loop diagram.
The lower cross indicates the chirality flip of $\chi$.
The result (\ref{numassB}) is obtained by using the exact propagators of $\phi$s and $\chi$s.
}
\end{figure}
\end{center}

The two-loop diagram for the neutrino mass is shown 
in Fig.~\ref{twoloop2}.
Because of  the soft breaking of the dimension two operator
$\phi^2$,
the propagator between $\phi$ and $\phi$ 
can exist, generating the mass of $N$: 
\begin{align}
(M_N)_{ij} 
&=\frac{1}{32 \pi^2}
\sum_k(Y^\chi_{ik})^*
(Y^\chi_{jk})^* M_{\chi_k}
\left[
\frac{m_{\phi_{\phi_R}}^2}{m_{\phi_{\phi_R}}^2-M_{\chi_k}^2}
\ln\left(\frac{m_{\phi_{\phi_R}}}{M_{\chi_k}}\right)^2  
-\frac{m_{\phi_I}^2}{m_{\phi_I}^2-M_{\chi_k}^2}
\ln\left(\frac{m_{\phi_I}}{M_{\chi_k}}\right)^2
\right]~.
\label{MR}
\end{align}
The 3$\times$3 two-loop neutrino mass matrix ${\cal M}_\nu$
is given by
\begin{align}
({\cal M}_\nu)_{ij} 
&= 
\frac{1}{(32\pi^2 )^2}
\sum_{l,k}  Y^\nu_{il}
Y^\nu_{jm} (Y^\chi_{lk})^*
(Y^\chi_{mk})^* M_{\chi_k}
(m_{\eta_I^0}^2-m_{\eta_R^0}^2)\nn\\
& \times \int_0^\infty dx ~
\frac{1}{(x+m_{\eta_R^0}^2)^2(x+m_{\eta_I^0}^2)}
 \int_0^1 dz \ln\left[
 \frac{z M_{\chi_k}^2+(1-z)m_{\phi_I}^2+z(1-z) x}{z M_{\chi_k}^2+(1-z)m_{\phi_R}^2+z(1-z) x}
 \right].
\label{numassB}
\end{align}
We can also use (\ref{MR}) to obtain an approximate formula
for the neutrino mass
\begin{align}
({\cal M}_\nu)_{ij} 
&\sim \frac{1}{32 \pi^2}
\sum_k Y'^\nu_{ik} 
Y'^\nu_{jk} M_k
\ln\frac{m_{\eta_R^0}^2}{m_{\eta_I^0}^2}
~,~
Y'^\nu_{jk} = Y^\nu_{jl} U^N_{lk},
\label{Mnu}
\end{align}
where $U^N$ is the unitary matrix diagonalizing
the mass matrix $(M_N)_{ij}$
with the eigenvalues $M_{k}$ and the mass eigenstates
$N'_k$,  and we have used the fact that $M_{k} 
\ll m_{\eta_R^0} \simeq m_{\eta_I^0}$.
In the following discussions we choose the theory parameters
so as to be consistent with
the global fit \cite{Esteban:2016qun}:
\be
\Delta m_{21}^2 &=& 7.50^{+0.19}_{-0.17}\times 10^{-5}
~\mbox{eV}^2,~\nn\\
\Delta m_{31}^2 &=&2.524^{+0.039}_{-0.040}
~(-2.514^{+0.038}_{-0.041})\times 10^{-3}\mbox{ eV}^2~,\nn\\\sin^2\theta_{12}
&=&0.306\pm 0.012,~
\sin^2\theta_{23}=0.441^{+0.027}_{-0.021}
~(0.587^{+0.020}_{-0.024}),~\label{Vmns}\\
\sin^2\theta_{13}&=&0.02166\pm 0.00075
~(0.02179\pm 0.00076),\nn
\ee 
where the values in the parenthesis
are those for  the inverted mass hierarchy.  

\subsection{Dark radiation}
According to the discussion at the beginning of this section,
we identify the lightest right-handed neutrino
with  dark radiation contributing to $\Delta N_{\rm eff}$
\footnote{Within a similar  framework of radiative seesaw mechanism, 
the lightest right-handed neutrino has been regarded
as stable warm dark matter
in \cite{Sierra:2008wj}.
In the models proposed in \cite{Kajiyama:2013rla,Baek:2013fsa},
the topology of the two loop to generate the neutrino mass is basically the same as 
that of Fig.~\ref{twoloop2}.
But the matter content of our model is much simpler; we have only two additional
extra fields compared with the 
one-loop model of Ma~\cite{Ma:2006km},
while in these papers  five and four additional ones have to be introduced.
Apart from this difference, 
they have not considered the lightest right-handed
neutrino as dark radiation.
In \cite{Baek:2013fsa}, however, 
the Nambu-Goldstone boson associated with the spontaneous breaking
of U(1)$_{\rm B-L}$ is regarded as dark radiation.
}.
Without los  of generality we may assume it is $N'_1$
with mass $\lsim 0.24$ eV. The upper bound on the mass is obtained together with $3.10 < N_{\rm eff}< 3.42$
in Ref.~\cite{Feng:2017nss}. 
To simplify the
situation,  we require that 
the heavier right-handed neutrinos $N'_2$ and $N'_3$
decay above the decoupling temperature $T_N^{\rm dec}$
of $N'_1$.
Their decay widths are given by
\begin{align}
 \langle~\Gamma(N'_{2(3)}\to N'_1\nu \bar{\nu}) +
 \Gamma(N'_{2(3)}\to \bar{N'}_1\nu \bar{\nu})~\rangle
 =
 \frac{1}{3072\pi^3}\frac{M_{2(3)}^5}{m_{\eta^0}^4}~\sum_{i,j}
 |Y'^{\nu}_{i2(3)}|^2|Y'^{\nu}_{j1}|^2~,
 \label{width}
\end{align}
where we have used 
 $m_{\eta^0}=m_{\eta_R^0}\simeq m_{\eta_I^0}$ and neglected the 
mass of $N'_1$ and $\nu_L$s.
Therefore,   $N'_2$ and $N'_3$ can decay above $T_N^{\rm dec}$
if
\begin{align}
 \langle~ \Gamma(N'_{2(3)}\to N'_1\nu \bar{\nu})+
 \Gamma(N'_{2(3)}\to \bar{N'}_1\nu \bar{\nu}) ~\rangle
 &\gsim   H(T_N^{\rm dec})
  \label{conditionH}
 \end{align}
 is satisfied, where $H(T)$ is the Hubble constant at temperature $T$,
 and $g_{*s}(T)$ is the relativistic degrees of freedom at $T$.
To obtain the effective
number of the light relativistic species $N_{\rm eff}$ 
\cite{Kolb:1990vq,Steigman:2012ve},
we have to compute the 
energy density of $N'_1$ at the time of 
the photon decoupling, where we denote
the decoupling temperature of $\gamma, \nu_L$ and $N'_1$
by $T_{\gamma 0}, T_\nu^{\rm dec}$ and $T_N^{\rm dec}$,
respectively. Further,
$T_{\nu 0}~(T_{N 0})$ stands for the temperature of
$\nu_L~(N'_1)$ 
at the decoupling of $\gamma$.
The most important fact is that the entropy per comoving volume is
conserved, so that $s a^3$ is constant, where
$s$ is the entropy density, and $a$ is the scale factor.
The effective number $N_{\rm eff}$   follows from
$\rho_r(T_{\gamma 0}) 
=(\pi^2/15)
(1+(7/8)( 4/11)^{4/3}
N_{\rm eff}) ~T_{\gamma 0}^4$
and is given by \cite{Kolb:1990vq,Steigman:2012ve,Anchordoqui:2011nh}
\be
N_{\rm eff} = 3.046+\left(\frac{g_{*s}(T_\nu^{\rm dec})}{g_{*s}(T_N^{\rm dec})}\right)^{4/3}
\label{Neff}
\ee
for $N_\nu=3$, where $\rho_r$ is the energy density of
relativistic species.
Since 
$g_{*s}(T_\nu^{\rm dec})=(11/2)+(7/4) N_\nu =10.75$, we need to compute
the decoupling temperature $T_N^{\rm dec}$
to obtain $g_{*s}(T_N^{\rm dec})$
and hence $ N_{\rm eff}$.
For $0.05 \lsim \Delta N_{\rm eff}\lsim 0.38$
\cite{Feng:2017nss}
we find
$101\gsim g_{*s}(T_N^{\rm dec})\gsim 22$
and also
$T_N^{\rm dec} \simeq 165$  MeV
to obtain $g_{*s}(T_N^{\rm dec}) \simeq 30$
(which gives $\Delta N_{\rm eff}=0.25$).
To estimate $T_N^{\rm dec}$, we  compute
the annihilation rate $\Gamma_N(T) $ of $N'_1$ at $T$, 
which is given by
\begin{align}
\Gamma_N(T) &=
n_N(T)~\left[  \langle\sigma_{N'_1 N'_1  \to  \nu_L\nu_L} v
 \rangle(T)
+\langle\sigma_{N'_1 \bar{N}'_1  \to  \nu_L\bar{\nu}_L} v 
 \rangle(T)\right]\nn\\
&=
\frac{\pi^5}{9\zeta(3)}\left(\frac{7}{120}\right)^2 
\sum_{i,j}|Y'^{\nu}_{i 1}|^2 |Y'^{\nu}_{j 1}|^2
\frac{T^5}{(m_{\eta^0})^4}~,
\label{GammaN}
\end{align}
where $\zeta(3)\simeq 1.202\dots$ and $n_N(T)$ is the number density of $N'_1$.
Then we calculate  $T_N^{\rm dec}$ 
from  $\Gamma_N(T_N^{\rm dec})= H(T_N^{\rm dec})$,
which can be rewritten as
\footnote{We use the relation between $ T$ and $ g_{*s}$ 
given in \cite{Husdal:2016haj}  to solve Eq.~(\ref{TN}) for $T_N^{\rm dec}$.}
\be
\left( \frac{T_N^{\rm dec}}{164.2~\mbox{MeV}} \right)^3
\left(   \frac{29.9}{g_{*s}(T_N^{\rm dec})} \right)^{\frac{1}{2}}
 &=& 
\left(   \frac{m_{\eta^0}}{200~\mbox{GeV}} 
\frac{0.0409}{Y^\nu}  \right)^{4}
\label{TN}
\ee
with
$(Y^{\nu})^2=\sum_{i}|Y'^{\nu}_{i 1}| ^2$.

It turns out that 
$M_{2,3}\sim {\cal O}(10)$ GeV
 to obtain
$\Delta N_{\rm eff}\sim 0.25$
while satisfying  $M_{1}\lsim 0.24$ eV.
To see this, we first find 
that 
\be
\left(\frac{m_{\eta^0}^2}{\sum_{i}|Y'^{\nu}_{i 1}| ^2} \right)
\sim 2.4  \times 10^7~\mbox{GeV}^2~,
\ee
which follows from (\ref{TN}) for $\Delta N_{\rm eff} \sim 0.25$.
Further we can estimate a part of (\ref{width})
 from the neutrino mass (\ref{Mnu})
with $M_\nu \sim 0.05$ eV:
\be
\frac{M_{2(3)}}{m_{\eta^0}^2}~\sum_{i,j}
 |Y'^{\nu}_{i2(3)}|^2\sim
 2.7\times 10^{-15} |\lambda_5|^{-1}~\mbox{GeV}^{-1}~,
\ee
where we have used $m^2_{\eta_R^0}-m^2_{\eta_I^0}\simeq \lambda_5 v_h^2$. Then using (\ref{conditionH}) with $T\simeq 165~
\mbox{MeV}$
(which corresponds to $\Delta N_{\rm eff}\simeq 0.25)$, we obtain
\be
M_{2, 3} \lsim 17|\lambda_5|^{1/4}~\mbox{GeV}~.
\ee
Note that this is an  order of magnitude estimate,
and indeed $M_{2(3)}$ can not be  smaller than
$10$ GeV to satisfy $\Delta N_{\rm eff}\lsim 0.38$.

Since we require that $M_{1} \lsim 0.24$ eV,
there exists a huge hierarchy in the right-handed neutrino mass.
This has a consequence on the Yukawa coupling matrix $Y'^\nu$:
To obtain  realistic neutrino masses with the mixing parameters
given in (\ref{Vmns}), 
\be
|Y'^\nu_{i1}| & \gg & |Y'^\nu_{i2(3)}|
\label{hierarchy}
\ee
has to be satisfied. Note that  only $|Y'^\nu_{i1}|$ enters
into the thermally averaged annihilation cross section
of $N'_1$,  as we can see from (\ref{GammaN}). 
Because of $\Delta N_{\rm eff}\lsim 0.38$, on the other hand,
$|Y'^\nu_{i1}|$ can not be made arbitrarily large.
The hierarchy (\ref{hierarchy}) 
has effects on the LFV radiative decays of the
type $l_i\to l_j \gamma$, so that
the LFV  decays and $\Delta N_{\rm eff}$ are related, as we will see below.
In the limit $m_j \ll m_i$, where $m_i$ and $m_j$ stand for the mass of 
$l_i$ and $l_j$, respectively, the 
ratio of the partial  decay width
$    \hat{B}(l_i  \to  l_j\gamma)=\Gamma(l_i \to  l_j\gamma)/ 
 \Gamma(l_i \to  \nu_i e \bar{\nu}_e)$
can be
written as \cite{Ma:2001mr}
\be
\hat{B}(l_i \to l_j \gamma) &=& 
  \left(
    \frac{\alpha  }{768\pi G_F^2}
  \right) 
  \frac{\left| \sum_k(Y'^{\nu}_{i k})^* Y'^\nu _{j k}
  \right|^2}{m_{\eta^\pm}^4}.
  \label{Bhat}
\ee
Here $m_{\eta^\pm}$ and $Y'^\nu_{ik}$ are defined in 
(\ref{massceta}) and (\ref{Mnu}), respectively, and
the current upper bounds on the branching fraction of these processes
\cite{TheMEG:2016wtm,Olive:2016xmw}  require
\begin{align}
&\mu \to  e\gamma:  \label{fcnc1}
\left|\sum_k (Y'^{\nu}_{2 k})^* Y'^\nu _{1 k}\right|  \lsim  2.5\times 10^{-4}
\left(\frac{m_{\eta^\pm}}{220\,\mbox{GeV}}\right)^2,\\
&  \tau 
\to  \mu\gamma:  \label{fcnc2}
\left|\sum_k (Y'^{\nu}_{3 k})^* Y'^\nu _{2 k}\right|  \lsim  8.1\times  10^{-2}
\left(\frac{m_{\eta^\pm}}{220\,\mbox{GeV}}\right)^2,\\
 & \tau  \to   e\gamma:  \label{fcnc3}
\left|\sum_k (Y'^{\nu}_{3 k})^* Y'^\nu _{1 k}\right|  \lsim  7.0\times   10^{-2}
\left(\frac{m_{\eta^\pm}}{220\,\mbox{GeV}}\right)^2.
\end{align}
From (\ref{fcnc1}) we find that $Y'^\nu _{31}$
is not constrained by  the stringent constraint
from $\mu\to e \gamma$,
which will be crucial in obtaining a realistic  $N_{\rm eff}$
without having any contradiction with (\ref{fcnc1})-(\ref{fcnc3}).
Furthermore, if  $Y'^\nu _{31}$ is large compared with others
and the hierarchy (\ref{hierarchy}) is satisfied,
the ratio
$R=\hat{B}(\tau\to \mu \gamma) \hat{B}(\tau\to e \gamma)/
 \hat{B}(\mu\to e \gamma)$ is $\sim |Y'^\nu _{31}|^2$,
 and from the same reason
$\Delta N_{\rm eff}$ depends mostly on $Y'^\nu _{31}$.
A benchmark set of the input parameters is given by
\begin{align}
Y'^{\nu}_{i j}&=
\left(
\begin{array}{ccc}
-0.0382&  2.510\times 10^{-5}&  3.349 \times 10^{-5}   \\
 0.00129 & -1.183\times 10^{-6} & 1.081 \times 10^{-4}\\
 0.0154 & -7.723\times 10^{-5} & 9.334 \times 10^{-5} 
\end{array}
\right), \\
M_1&=0.147 {\rm ~eV}, ~~~M_2=M_3=9.55 {\rm ~GeV}, \\
m_{\eta^\pm} &=220  {\rm ~GeV}, ~~~m_{\eta_R^0}=200  {\rm ~GeV}, ~~~m_{\eta_I^0}= 207 {\rm ~GeV},
\end{align}
which yields
\begin{align}
\sin^2\theta_{12}&=0.305, ~~ \sin^2\theta_{23}=0.441, ~~ \sin^2\theta_{13}=0.0213,   \\
\Delta m_{21}^2 &=7.50 \times10^{-5} {\rm ~GeV}^2, ~~
\Delta m_{31}^2=0.00248 {\rm ~GeV}^2,\\
\hat{B}(\mu\to e \gamma)&=2.30\times 10^{-14},~
\hat{B}(\tau\to \mu \gamma)=3.75\times 10^{-15},~
\hat{B}(\tau\to e \gamma)=3.31\times 10^{-12},
\end{align}
where we have assumed that $Y'^{\nu}_{i j}$ are all real so that
there is no CP phase. These values  are consistent
with (\ref{Vmns}), (\ref{fcnc1}) - (\ref{fcnc3}).
With the same input parameters we find:
The lhs of (\ref{conditionH})
=5.46$\times 10^{-21}$
 (1.78$\times10^{-20}$)~
GeV for $N_2~(N_3)$,
where the rhs is $H= 2.10\times 10^{-20}$ GeV
with
$
T^{\rm dec}_N =166.8 ~\mbox{MeV}
$
and
$g_{*s}(T^{\rm dec}_N )=30.83$, 
and
$\Delta N_{\rm eff} =0.245$.

\begin{center}
\begin{figure}
  \includegraphics[width=10cm]{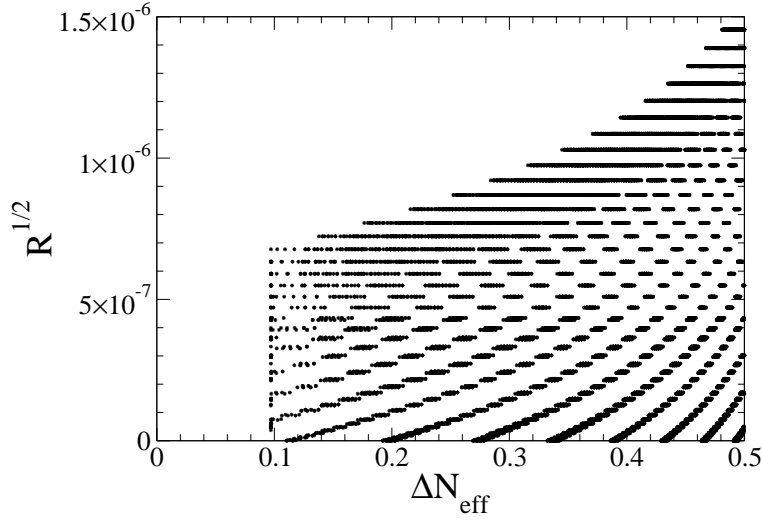}
\caption{\label{neff}\footnotesize
$R^{1/2}$ against 
$\Delta N_{\rm eff}$ with $m_{\eta^\pm}=240$ GeV and
$m_{\eta_R^0}=220$ GeV, where 
$m_{\eta_I^0}$ is varied between $221$ and $227$ GeV, and $R=
\hat{B}(\tau\to \mu \gamma) \hat{B}(\tau\to e \gamma)/
 \hat{B}(\mu\to e \gamma)$.
}
\end{figure}
\end{center}

 In Fig.~\ref{neff}  we plot $R^{1/2}$ against 
$\Delta N_{\rm eff}$ with $m_{\eta^\pm}=240$ GeV and
$m_{\eta_R^0}=220$ GeV, where 
we have varied $m_{\eta_I^0}$ between $221$ and $227$ GeV.
In the black region of Fig.~\ref{neff} the differences
of the neutrino mass squared and the neutrino  mixing angles 
are consistent with (\ref{Vmns}) for the normal hierarchy, and the constraints 
$M_1 < 0.24$ eV, (\ref{conditionH}) and (\ref{fcnc1})-(\ref{fcnc3})
are satisfied.
If $\Delta N_{\rm eff}$ and $R^{1/2}$  would depend on
$Y'^\nu _{31}$ only, we would obtain a line in the 
$\Delta N_{\rm eff}$ - $R^{1/2}$ plane.
The $Y'^\nu _{11}$ and $Y'^\nu _{21}$
dependence in $R^{1/2}$ cancels, but this is not the case for
$\Delta N_{\rm eff}$.
This is the reason why we have an area instead of a line
in Fig.~\ref{neff}.
We see from Fig.~\ref{neff} that 
the predicted region for  $\Delta N_{\rm eff}
\lsim 0.1$ is absent.
The main reason is that we have assumed that 
$M_2, M_3 \lsim 16$ GeV. This has also a consequence on the difference
between $m^2_{\eta^0_R}$ and $m^2_{\eta^0_I}$, because the mass difference
changes the overall scale of the neutrino mass (\ref{Mnu}).
To obtain a larger $M_{2,3}$, we can decrease the mass difference, 
thereby implying an increase of the degree of fine-tuning.
Further, the difference
between $m^2_{\eta^0_R}$ and $m^2_{\eta^0_I}$ can not be made
arbitrarily large,  because  it requires a smaller  $M_{2,3}$, which
due to $H(T) \propto T^2$  in turn implies that
the decoupling temperature $T_N^{\rm dec}$ has to decrease
to satisfy the constraint (\ref{conditionH}). 
A smaller $T_N^{\rm dec}$, on the other hand,
means a larger $\Delta N_{\rm eff }$ which is constrained to be below
$0.38$. This is why $m_{\eta_I^0}$ is varied only in a small interval
in Fig.~\ref{neff}.
Since the current upper bound
on $ B(\mu\to e \gamma) \simeq\hat{B}(\mu\to e \gamma)$
is $4.2\times 10^{-13}$~\cite{TheMEG:2016wtm}, the model B predicts
\begin{align}
& \left[ B(\tau\to \mu \gamma) B(\tau\to e \gamma)\right]^{1/2}
\simeq
\left[ \frac{\hat{B}(\tau\to \mu \gamma)}{0.17}~
\frac{\hat{B}(\tau\to e \gamma)}{0.18}\right]^{1/2}
\lsim 1.2 \times 10^{-10}~,
\end{align}
which is about two orders of magnitude smaller than the 
current experimental bounds \cite{Olive:2016xmw}.

\begin{center}
\begin{figure}[t]\vspace{1cm}
  \includegraphics[width=10cm]{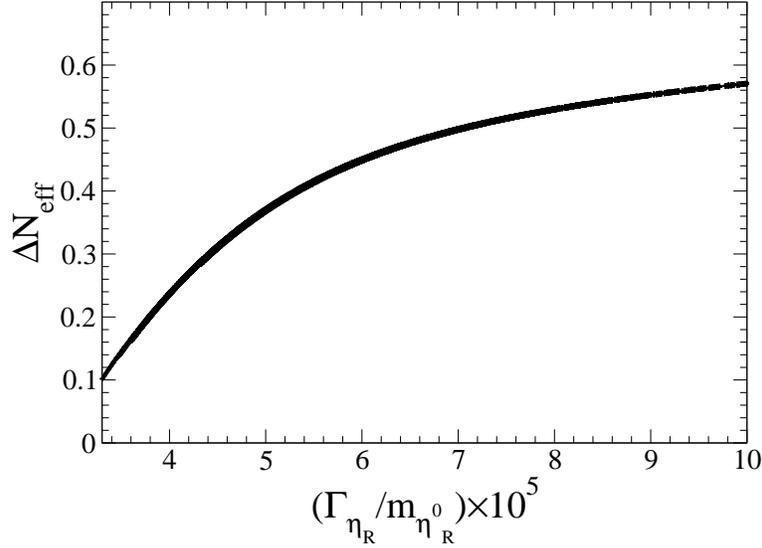}
\caption{\label{eta-decay}\footnotesize
$\Delta N_{\rm eff}$ against  $\Gamma_{\eta_R}/m_{\eta^0_R}$, 
where we have used the same parameters as for Fig.~\ref{neff}.
}
\end{figure}
\end{center}

Another consequence of the hierarchy (\ref{hierarchy}) is that
the total decay width of $\eta_R$ depends on 
$\sum_{i,j} |Y'_{ij}|^2$, which is
approximately $\sum_{i} |Y'_{i1}|^2$
(we assume that  $\eta_R$ is the lightest among $\eta$s).
Therefore, $\Delta N_{\rm eff}$ is basically a function
of the decay width. In Fig.~\ref{eta-decay} we show
$\Delta N_{\rm eff}$ against  $\Gamma_{\eta_R}/m_{\eta^0_R}$, 
the decay width of $\eta^0_R$ over $m_{\eta^0_R}$ , 
where we have used the same parameters as for Fig.~\ref{neff}.
$\eta^0_R$ decays almost 100 percent into neutrinos and dark radiation
$N'_1$, which is invisible. In contrast to this, $\eta^+$ can decay
into a charged lepton and $N'_1$, and the decay width over $m_{\eta^\pm}$
is the same as $\Gamma_{\eta_R}/m_{\eta^0_R}$.
$\Gamma_{\eta_R}$  should be compared with the  decay width for
$\eta^+ \to W^{+*} ~\eta^0_{R,I} \to \mbox{} f \bar{f'}~N'_1
\nu$,
which is 
$\sim 10^{-8}m_{\eta^\pm}$ 
for the same parameter space
as for Fig.~\ref{eta-decay}, where $f$ and $f'$ stand for the SM fermions (except the top quark). Therefore, $\eta^+$ decays almost 100 percent 
into a charged lepton and missing energy.
In Ref.~\cite{Sierra:2008wj}, a similar hierarchical spectrum of the right-handed neutrinos in the model of ~\cite{Ma:2006km} has been assumed (the lightest one has been
regarded as a warm dark matter) and  collider physics has been discussed.
How the inert Higgs bosons
can be produced via $s$-channel exchange 
of a virtual 
photon and $Z$ boson~\cite{Djouadi:2005gj}
 is the same, but the decay of the inert Higgs bosons is different because of the hierarchy
 (\ref{hierarchy}) of the Yukawa coupling constants.
As it is mentioned above, the $\eta^\pm$ decays in the present model almost only into
the lightest one $N'_1$ and a charged lepton.
Therefore, the cascade decay of the heavier 
right-handed neutrinos into  charged leptons will not be seen 
at collider experiments, because they can be produced  
only as a decay product of  $\eta^\pm$.
The decay width of $\eta^\pm$  into an individual charged lepton depends of course on
the value of $Y'^\nu_{i1}$. 
In the parameter space we have scanned we cannot 
make any definite conclusion on the difference.
\begin{center}
\begin{figure}[t]
\vspace{10mm}
  \includegraphics[width=12cm]{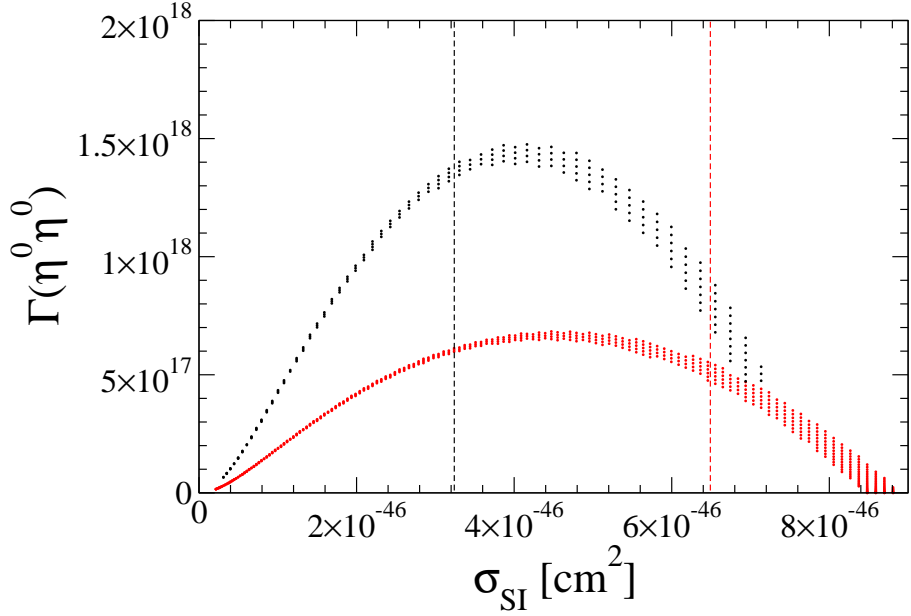}
\vspace{-5mm}
\caption{\label{rate}\footnotesize
The pair-annihilation rate of $\phi_R$ into $\eta^0_R\eta^0_R$
and $ \eta^0_I\eta^0_I$ in the Sun,
$\Gamma(\eta^0 \eta^0)=\Gamma(\phi_R\phi_R\to 
\eta^0 \eta^0;t_0)$,
against $\sigma_{\rm SI}$ for 
$m_{\phi_R} = 250$ (black) and $500$ (red) GeV, where 
$m_{\eta}$ is fixed at $230$ GeV (all $\eta$s have the same mass)
and $0.117 < \Omega_{\phi_R}h^2
< 0.123$. The black (red) vertical dashed line 
is the XENON1T \cite{Aprile:2017iyp} upper bound on $\sigma_{\rm SI}$
for $m_{\phi_R} = 250$ (black) and $500$ (red) GeV.
}
\end{figure}
\end{center}
\subsection{Cold dark matter and 
its direct and indirect detection}
Since the lightest $N$ is dark radiation and 
the masses of the heavier ones  are ${\cal O}(10)$ GeV
(as we have seen in the previous subsection),
$\eta^0_{R,I}$ can not be DM candidates,
because they decay into $N$ and $\nu$.
So, DM candidates are 
$\chi$ and the lightest component of $\phi$
\footnote{Both together can not be DM, because the heavier one
decays into $N'_1+\mbox{lighter one}$.}.
In the case that  $\eta$s are lighter than  $\phi_R$
and the lightest component of $\phi$ (which is assumed to be
$\phi_R$) is DM, 
a correct relic abundance
$\Omega_{\phi_R} h^2=0.1204\pm 0.0027$ 
\cite{Ade:2013zuv} can be easily obtained, 
because $\gamma_3$
for the scalar coupling $(\eta^\dag \eta) |\phi|^2  $
is an unconstrained parameter so far. 
So, in the following discussion we assume that 
$\phi_R$ is DM.

Because of the Higgs portal coupling $\gamma_2$,
the direct detection of $\phi_R$ is 
possible.
The current experimental bound of XENON1T \cite{Aprile:2017iyp} of
the spin-independent cross section $\sigma_{\rm SI}$  off the nucleon 
requires
$|\gamma_2 | 
\lsim 0.05\sim 0.14~ \mbox{for}~m_{\phi_R}
=250\sim 500~\mbox{GeV}$.
Since $\gamma_2$ is allowed only below an upper bound
(which depends on the DM mass $m_{\phi_R}$), $\gamma_3$
can vary in a certain interval for a given DM mass.

With this remark, we note that the capture rate
of DM in the Sun  is proportional to $\sigma_{\rm SI}$, while
its annihilation rate in the Sun is proportional 
 to the thermally averaged annihilation cross section,
$\langle v\sigma(\phi_R\phi_R\to \eta^+\eta^-, 
\eta^0_R\eta^0_R, \eta^0_I \eta^0_I ) \rangle$
\cite{ Silk:1985ax, Griest:1986yu,  Ritz:1987mh,   Kamionkowski:1991nj,  Jungman:1995df}.  
If a pair of $\phi_R$s annihilates
into $\eta^0_R\eta^0_R$ and also $\eta^0_I\eta^0_I$,
a pair of $\nu_L$ and $\bar{\nu}_L$
will be produced, which may be observed on the Earth
\cite{Agrawal:2008xz}.
The signals will look very similar to those coming
from $W^\pm$, which result from  DM annihilation.
The annihilation rate as a function of time $t$ is given by
\cite{Jungman:1995df}
\begin{align}
&\Gamma(\phi_R\phi_R\to\eta^0_R\eta^0_R, \eta^0_I \eta^0_I;t)
=\Gamma(\phi_R\phi_R\to\eta^0\eta^0;t)\nn\\
&=\frac{1}{2}
\frac{C_{\phi_R} C_A(\eta^0\eta^0)}{C_A(\eta^+\eta^-)+C_A(\eta^0\eta^0)+
C_A(XX')} 
\tanh^2 \left[t \sqrt{(C_A(\eta^+\eta^-)+C_A(\eta^0\eta^0)+
C_A(XX'))C_{\phi_R}}\right],
\end{align}
where $C_{\phi_R}$ is the capture rate in the Sun, 
\begin{eqnarray}
C_{\phi_R} &\simeq &
 1.4\times 10^{20} f(m_{\phi_R})\left( \frac{\hat{f}}{0.3} \right)^2 
\left( \frac{\gamma_{2}}{0.1} \right)^2 
\left( \frac{200~\mathrm{GeV}}{m_{\phi_R}} \right)^2
\left( \frac{125~\mathrm{GeV}}{m_{h}} \right)^4 ~,
\ee
and $C_A$ is given by 
\begin{align}
C_A(\bullet )
&=\left(\frac{\langle  \sigma_{\phi_R\phi_R\to\bullet}\, v\rangle }{5.7\times 10^{27}
\mbox{cm}^3}
\right)
\left(\frac{m_{\phi_R}}{100 ~\mbox{GeV}}\right)^{3/2}~\mbox{s}^{-1}
~\mbox{with}~\bullet=\eta^+\eta^-,~\eta^0\eta^0, ~\mbox{and}~XX'~.
\end{align}
We have used $f(250 ~\mbox{GeV})\simeq 0.5 $ 
and $f(500~\mbox{GeV})\simeq 0.2$ \cite{Jungman:1995df}, and we
have assumed that  all the $\eta$s have the same mass
and therefore $C_A(\eta^0\eta^0)=
C_A(\eta^+\eta^-)$.
In Fig. \ref{rate} we plot the annihilation rate 
$\Gamma(\phi_R\phi_R\to 
\eta^0\eta^0;t_0)$ today ($t_0=1.45\times 10^{17}$ s) against  $\sigma_{\rm SI}$
for $m_{\phi_R} = 250$ and $500$ GeV with
$m_{\eta}$ fixed at $230$ GeV and $0.117 < \Omega_{\phi_R}h^2
< 0.123$. The vertical dashed lines are the XENON1T upper bound
on $\sigma_{\rm SI}$ \cite{Aprile:2017iyp}.
The peak of $\Gamma(\phi_R\phi_R\to 
\eta^0 \eta^0;t_0)$ for $m_{\phi_R} = 250$ (500) GeV
 appears at  $\sigma_{\rm SI}=4.2
~(4.7) \times 10^{-46}$ cm$^2$ and is $\simeq 1.7~(0.7) \times
10^{18}~\mbox{s}^{-1}$,
which is two to three  orders of magnitude smaller  than
the upper bound on the DM annihilation rate into
$W^\pm$ in the Sun \cite{Aartsen:2016zhm}  .

\section{Conclusion}

We have discussed the extensions of the Ma model by imposing a larger unbroken symmetry  $Z_2 \times Z_2'$.
Thanks to the symmetry, at least two stable particles exit.
We have studied two models, model A and model B, where the stable particles form a multicomponent DM system
in the model A, while they are a DM and dark radiation in the model B. 

The model A is an extension of  the model of Ma such that
the lepton-number violating ``$\lambda_5$ coupling", which is
$\mathcal{O}(10^{-6})$ to obtain small neutrino masses for $Y^\nu \sim 0.01$,
is radiatively generated.
Consequently, the neutrino masses are generated at the two-loop level, where the unbroken $Z_2\times Z_2'$
symmetry acts to forbid the generation of the one-loop mass.
Such larger unbroken symmetry implies that the model involves a multi-component DM system.
We have considered the case of the three-component DM system:
two of them are SM singlet real scalars and 
the other one 
is a right-handed neutrino.
The DM conversion and semi-annihilation in addition to the standard annihilation 
are relevant to the DM annihilation processes.
We have found that the non-standard processes have a considerable influence on the DM relic abundance. 
We also have discussed the monochromatic neutrinos from the Sun as the indirect signal of the semi-annihilation 
of the DM particles. 
In the cases of one-component DM system of
a real scalar boson or of a Majorana fermion 
the monochromatic neutrino production by the 
 DM annihilation is 
 strongly suppressed due to the chirality of the left-handed neutrino.  
However, such suppression is absent 
when  DM is a complex scalar boson or a  Dirac fermion.
Also in a multicomponent DM system, the neutrino production is unsuppressed
if it is an allowed process.
We have found that the rate for the monochromatic neutrino production in the model A is very small 
compared with the current IceCUBE~\cite{Aartsen:2016zhm} sensitivity. However, the resonant effect in the $s$-channel process
of the semi-annihilation can be expected to enhance the rate. 

In the model B, the mass of the right-handed neutrinos are produced at the one-loop level.
Then the radiative seesaw mechanism works at the two-loop level.
Thanks to $Z_2\times Z'_2$ there exist at least two stable DM particles;
a dark radiation $N'_1$ 
with a mass of ${\cal O}(1)$ eV and the other one, DM,
is the real part  of $\phi$.
The dark radiation contributes to  $\Delta N_{\rm eff } < 1$
such that the tensions in cosmology that exist among the observations in the local Universe
(CMB temperature fluctuations and primordial gravitational fluctuations)
can be alleviated.
Because of the hierarchy $M_{2,3} \gg T^{\rm dec}_N \simeq {\cal O}(100)~ \mbox{MeV}
\gg  M_{N_1}~{\cal O}(1)$ eV, we are able to relate to the ratio of 
the lepton flavor violating decays to $\Delta N_{\rm eff}$.
The indirect signal of the neutrino from the Sun has also been discussed. It is found that the predicted 
annihilation rate of the neutrinos is two to three orders of magnitude smaller than the current bound~\cite{Aartsen:2016zhm}.
We have also expressed $\Delta N_{\rm eff}$ as a function of the decay width of $\eta_R^0$ (which is assumed to be lightest among $\eta$s). It decays 100 percent into left- and right-handed neutrinos,  where the heavier right-handed neutrinos decay further into dark radiation (the lightest among them). 
Dark radiation appears as a missing energy
in collider experiments.
We also have found  that $\eta^+$  decays 100 percent
into a charged lepton and the missing energy.
This is a good example in which, through the generation mechanism 
of the neutrino masses, cosmology and collider physics are closely related.

\vspace*{5mm}

\subsection*{Acknowledgements}
The work of M.~A. is supported in part by the Japan Society for the
Promotion of Sciences Grant-in-Aid for Scientific Research (Grant
No. 25400250 and No. 16H00864). 
J.~K.~is partially supported by the Grant-in-Aid for Scientific Research (C) from the Japan Society for Promotion of Science (Grant No.16K05315).

\end{document}